\documentclass[
reprint, 
 superscriptaddress,
 amsmath,amssymb,
 aps,
]{revtex4-2}

\usepackage{graphicx}
\usepackage{dcolumn}
\usepackage{bm}

\usepackage{color}
\usepackage[colorlinks=true,citecolor=blue,linkcolor=blue,filecolor=blue, urlcolor=blue]{hyperref}

\begin{document}

\preprint{APS/123-QED}

\title{Competing interlayer interactions in twisted monolayer-bilayer graphene: \\ From spontaneous electric polarization to quasi-magic angle}


\author{Wei-En Tseng}
\affiliation{Institute of Atomic and Molecular Sciences, Academia Sinica, Taipei 10617, Taiwan}
\affiliation{Department of Physics, National Taiwan University, Taipei 10617, Taiwan}

\author{Mei-Yin Chou}
\email[Contact author: ]{mychou6@gate.sinica.edu.tw}
\affiliation{Institute of Atomic and Molecular Sciences, Academia Sinica, Taipei 10617, Taiwan}
\affiliation{Department of Physics, National Taiwan University, Taipei 10617, Taiwan}


\date{\today}

\begin{abstract}

The family of moiré materials provides a powerful platform for tuning interlayer couplings via the twist angle in systems with large spatial periodicity. In trilayer graphene systems, interlayer couplings at the two interfaces can possibly be tuned separately, and the competition between these interactions can therefore influence the electronic structure in a significant way. In this study, we investigate the electronic properties of twisted monolayer-bilayer graphene (\textit{aAB}) beyond the continuum model, using first-principles calculations combined with an accurate tight-binding model. We find that at large twist angles, the electronic features of \textit{aAB} are well described by the interaction between the parabolic bands of the Bernal \textit{AB} bilayer and the Dirac bands of the twisted monolayer \textit{a}, resulting in a spontaneous electric polarization in the former that splits the parabolic bands. As the twist angle decreases, the coupling between adjacent layers at the twisted interface becomes dominant, which makes \textit{aAB} look like twisted bilayer graphene (TBG) interacting with the outer Bernal layer \textit{B}. A moiré potential emerges in the TBG-like layers, leading to charge localization, while the outer Bernal layer \textit{B} exhibits charge delocalization with substantial sublattice polarization at the atomic scale. Furthermore, we identify narrow bands with a minimum width at a quasi-magic angle of $\theta=1.16^\circ$, closely matching the magic angle of TBG. The enhanced electron correlation expected in these narrow bands suggests that \textit{aAB} is a promising platform for exploring correlated electronic phenomena.

\end{abstract}

\maketitle

\section{Introduction}

The strong electron correlation in twisted bilayer graphene (TBG) has given rise to unexpected electronic properties such as correlated insulating phases \cite{correlated, correlated+sc+mag2}, superconductivity \cite{correlated+sc+mag2, sc+TBG}, orbital ferromagnetism \cite{mag1,correlated+sc+mag2}, and Chern insulators \cite{Chern1,Chern2,Chern3}. 
These exotic phenomena are believed to emerge from the formed flat bands, where the kinetic energy is reduced and the electronic correlation energy is enhanced. For TBG, the flat bands near the Fermi level with an almost zero velocity at K are particularly special as they only occur at a series of ``magic" twist angles between the two graphene layers, with the first magic angle at about $\theta \approx 1.1^\circ$ \cite{Bistritzer2011}. It has been proposed that these magic angles arise from matching the moiré periodicity with the size of the Fermi ring of \textit{AA} bilayer graphene in momentum space \cite{PhysRevB.110.115154}. 

Since the discovery of TBG, efforts have been made to search for flat bands in other twisted graphene systems. For example, twisted trilayer graphene (TTG) with different geometric stacking patterns has been investigated, including alternating twisted trilayer graphene (\textit{a}-TTG) \cite{Stephen_2020}, helical trilayer graphene (hTG) \cite{Devakul2023}, and twisted monolayer-bilayer graphene (\textit{aAB}) \cite{theory2}. The various stacking possibilities make TTG a more versatile platform than TBG.  Among them, twisted monolayer-bilayer graphene is the easiest to grow in experiments. It can be constructed by twisting monolayer graphene on top of a Bernal bilayer, resulting in a moiré pattern as shown in Fig. \ref{fig:moire}. We denote this system by \textit{aAB}, where \textit{a} represents the twisted monolayer, and \textit{AB} represents the untwisted underlying Bernal bilayer.

In a special range of twist angles $\theta \approx 1.0^\circ$\text{-}$1.4^\circ$, the \textit{aAB} system exhibited strong electrical tunability of correlated insulating states \cite{Shuigang_2021,  H.Polshyn_2020, Canxun_2023, H.Polshyn_2022, Shaowen_2021, Minhao_2021, Si_2022} and van Hove singularities \cite{Shuigang_2021}. When an external electric field was applied, the quantum anomalous Hall effect in an orbital Chern insulator was realized at a filling of three electrons per moiré unit cell \cite{H.Polshyn_2020,Canxun_2023}. The Chern number and associated magnetic order could further be switched by tuning the gate voltage, making \textit{aAB} a possible system to realize non-volatile switching of magnetization \cite{H.Polshyn_2020}. For other integer fillings, abundant orbital magnetism with the anomalous Hall effect was also reported within the correlated phases by transport measurements \cite{Minhao_2021, Shaowen_2021}. At fractional fillings, a Chern insulator with topological charge density waves has been reported to break the translational symmetry of the moiré superlattice \cite{H.Polshyn_2022}. These fascinating phenomena emerge from the interplay between electron correlation and nontrivial band topology in the \textit{aAB} system \cite{Shuigang_2021, Minhao_2021,Canxun_2023, Si_2022}. On the other hand, angle-resolved photoemission spectroscopy measurements have been performed to image the electronic structure for twist angles of $\theta=2^\circ$\text{-}$4^\circ$, providing evidence for electrically tunable parabolic bands and Dirac bands \cite{Nunn_2023,zhang2024}. 

The rich phase diagram observed at small twist angles has inspired extensive theoretical studies to unveil the electronic structure, focusing on the flat-band topology \cite{Liu2019, theory1, theory2, theory3} and the correlated insulating states \cite{ZhangShihao2022, theory3}. Due to the large supercell size with a moiré lattice constant $L \propto \theta^{-1}$ for small $\theta$, first-principles calculations become unfeasible and theoretical studies have predominantly relied on continuum models \cite{theory1,theory2, theory3, Liu2019} to date. Intuitively, the main feature to be considered in the \textit{aAB} trilayer should be the interaction between the parabolic bands of the \textit{AB} bilayer and the Dirac bands of the twisted monolayer \textit{a} \cite{Nunn_2023,zhang2024,theory1}. However, as we will show below, this picture breaks down at sufficiently small twist angles, as the interlayer coupling between layers \textit{a} and \textit{A} has a significantly stronger effect than the interaction within the \textit{AB} bilayer. The resulting trilayer system can be viewed as a TBG (\textit{a}+\textit{A}) interacting with the outer Bernal layer (\textit{B}). Therefore, the evolution and competition of interlayer interactions in the \textit{aAB} trilayer should play a crucial role in determining the electronic properties. Moreover, interlayer interactions should be carefully considered, as even the second-nearest-neighbor interlayer hopping terms within the \textit{AB} bilayer can significantly influence the band structure \cite{theory1,theory2}.

In this work, we perform first-principles calculations within density functional theory (DFT) and calculations using a tight-binding (TB) model with environment-dependent interlayer couplings \cite{TB_parameter_1} in order to provide an accurate description of the electronic properties of \textit{aAB} beyond the simplified continuum models. Our results indicate that two distinct regimes exist as a result of the evolution of interlayer interactions as the twist angle varies.  At large twist angles, \textit{aAB} behaves as an \textit{AB} bilayer interacting with a twisted monolayer \textit{a}, with a 30-40 meV splitting of the parabolic bands resulting from the spontaneous electric polarization induced by the broken crystal symmetry. As the twist angle decreases below $1.3^\circ$, the \textit{aAB} system is better described as a TBG interacting with the outer Bernal layer, because the interlayer interaction between layers \textit{a} and \textit{A} has a dominant effect over the interaction within the \textit{AB} bilayer. Consequently, \textit{aAB} exhibits a behavior similar to that of TBG, with low-energy bands reaching a minimum width at an angle similar to the first magic angle of TBG, which we identify as the quasi-magic angle (qMA). In addition, the narrow bands near the qMA exhibit charge localization in the TBG-like layers and charge delocalization in the outer Bernal layer with substantial sublattice polarization.

\section{Computational Details}
\begin{figure}[htb]
    \centering
    \includegraphics[width=0.45\textwidth]{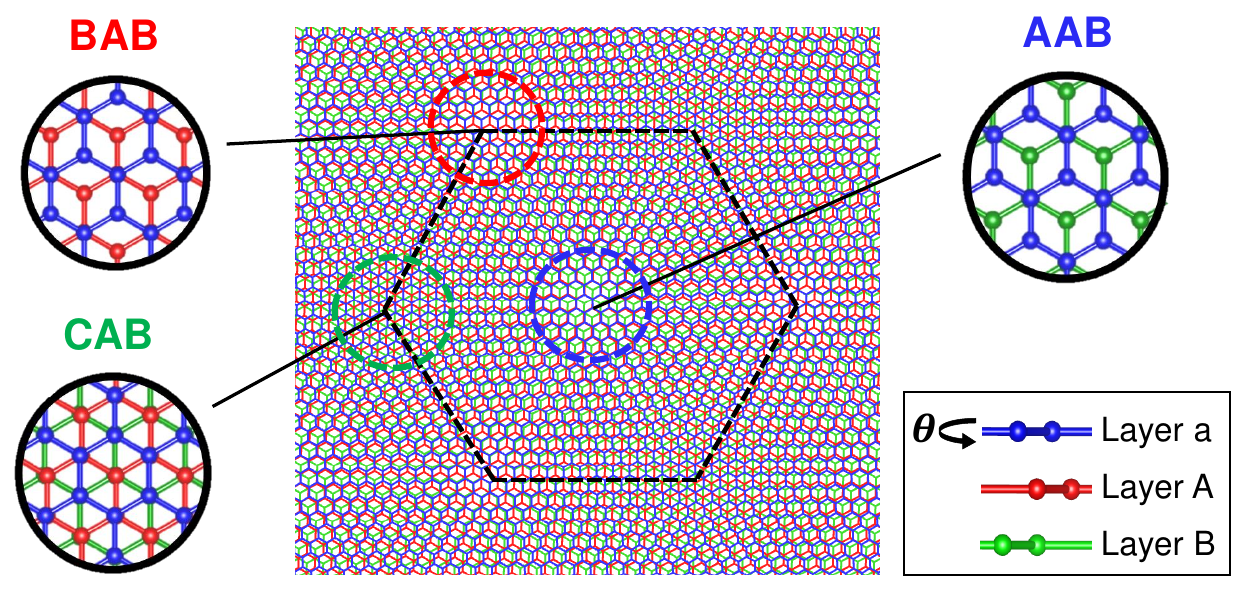}
    \caption{Moiré pattern of \textit{aAB} with a twist angle of $\theta=2.64^\circ$. The top layer (labeled as layer \textit{a}) is rotated with respect to the \textit{AB}-stacked bilayer. In our convention, layers \textit{A} and \textit{B} refer to the middle and bottom layers, respectively. The black dashed hexagon represents the Wigner-Seitz moiré supercell, with the three high-symmetry stacking patterns labeled as \textit{AAB}, \textit{BAB}, and \textit{CAB}. }
    \label{fig:moire}
\end{figure}

In this work, we consider commensurate \textit{aAB} structures with the moiré lattice vectors defined as ${\textbf{L}_1}=m{\textbf{a}_1}+(m+1){\textbf{a}_2}$ and $\textbf{L}_2=-(m+1){\textbf{a}_1}+(2m+1){\textbf{a}_2}$, where $m$ is an integer, and ${\textbf{a}_1}$ and ${\textbf{a}_2}$ represent the lattice vectors of monolayer graphene. The twist angle can be determined by $\theta=\cos^{-1}\left[(3m^2+3m+\frac{1}{2})/(3m^2+3m+1)\right]$.
In Fig. \ref{fig:moire}, we illustrate the moiré pattern of \textit{aAB} with a twist angle of $\theta=2.64^\circ\;(m=12)$. The local atomic configurations vary within the moiré supercell, and the three high-symmetry stacking patterns are labeled as \textit{AAB}, \textit{BAB}, and \textit{CAB}, respectively. 

The electronic structure of \textit{aAB} at large twist angles is studied using DFT calculations. We use the Vienna Ab initio Simulation Package (VASP) \cite{Kresse1996,Kresse1996-2} with the projector-augmented wave method \cite{Blöchl1994, Joubert1999} and the exchange-correlation functional in the generalized gradient approximation (GGA) \cite{Perdew1996}. The plane-wave cut-off energy is 500 eV. We use a k-point mesh of $12\times12\times1$, $9\times9\times1$, $6\times6\times1$ and $3\times3\times1$ centered at $\Gamma$ for \textit{aAB} supercells with $\theta=13.2^\circ$ ($m$=2), $\theta=9.4^\circ$ ($m$=3), $\theta=5.1^\circ$ ($m$=6), and $\theta=3.1^\circ$ ($m$=10), respectively. In addition, the Berry phase method \cite{King-Smith1993} is used to calculate the dipole moment of the supercell with a denser k-point mesh: $18\times18\times1$ for $\theta=13.2^\circ$ ($m$=2). In our calculations, the interlayer distances between adjacent layers are fixed at $d=3.35$ \AA, and a vacuum of 15 \AA\; is used to prevent interactions between artificial periodic cells. 

An accurate TB model with environment-dependent parameters \cite{TB_parameter_1} is used to investigate the electronic properties of \textit{aAB} for a wide range of twist angles. In our TB calculations, we take into account eight nearest-neighbor intralayer couplings. For interlayer couplings, we include the hoppings when the projected in-plane distance between two atoms is smaller than the fifth nearest-neighbor distance. The integrated charge distribution for the four low-energy flat bands was calculated using a $\Gamma$-centered k-point mesh of $18\times18\times1$. The interlayer distances between adjacent layers are also set to $d=3.35$ \AA. 

\section{Results and Discussions}
\subsection{Electric polarization induced by interlayer coupling }

\begin{figure}[htb]
    \centering
    \includegraphics[width=0.48\textwidth]{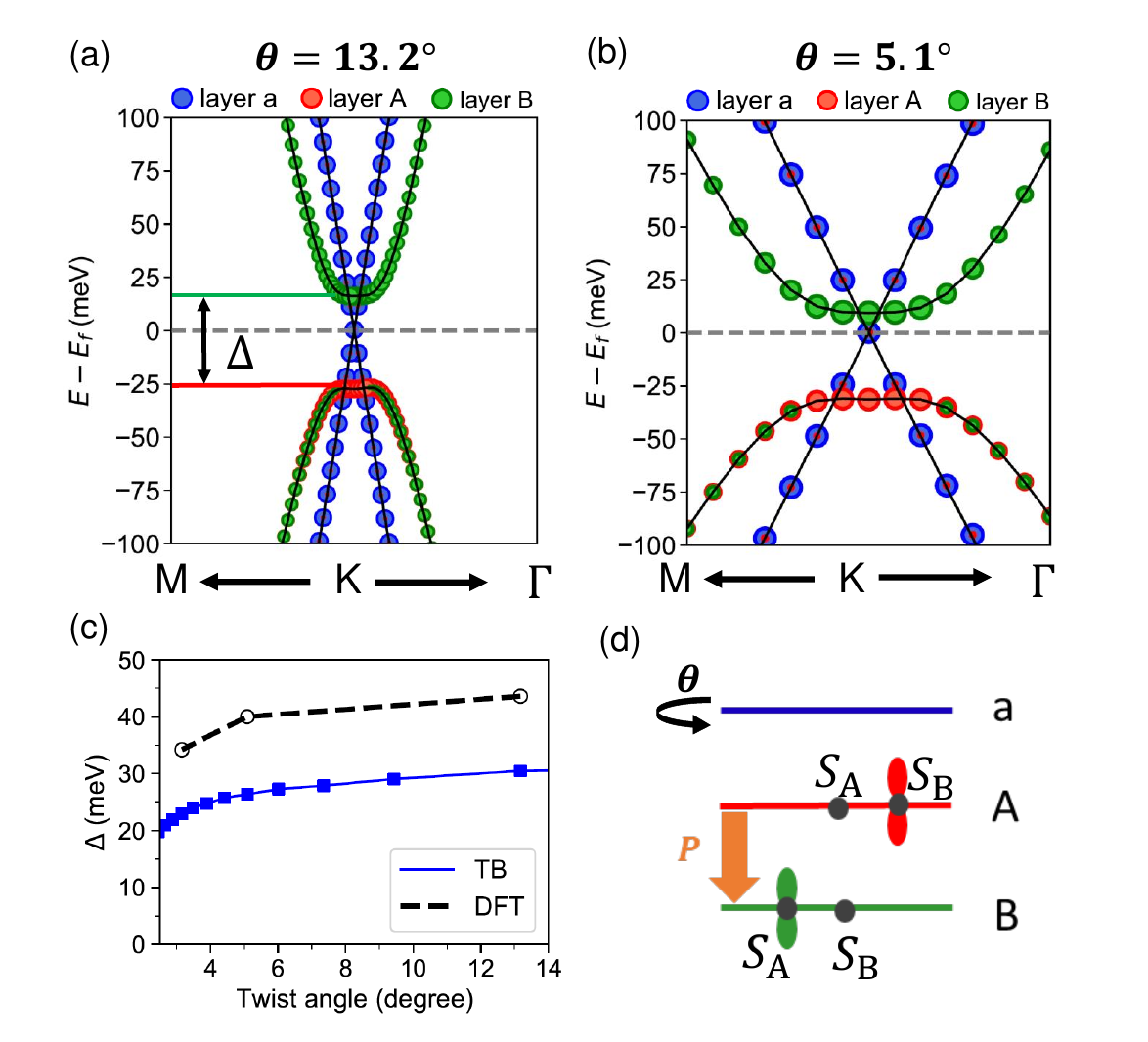}
    \caption{Layer-projected band structure of \textit{aAB} near the K point for twist angles (a) $\theta=13.2^\circ$ and (b) $\theta=5.1^\circ$ obtained by DFT calculations. At these large twist angles, the parabolic bands from the \textit{AB} bilayer (red and green) have a significant splitting $\Delta$ of about 40 meV induced by a spontaneous electric polarization (see text). The Dirac bands are mainly composed of layer \textit{a} states (blue).  The lengths along $\protect\overrightarrow{KM}$  and $\protect\overrightarrow{K\Gamma}$ directions are taken as a half of the distance from K to M.  (c) Comparison of the splitting $\Delta$ calculated by DFT and the TB model as a function of the twist angle. (d) Spontaneous electric polarization in \textit{aAB} caused by layer-asymmetric charge distribution. The red and green $p_z$ orbitals show the layer- and sublattice-dependent states at the edge of the parabolic valence and conduction bands, respectively. }
    \label{fig:proband}
\end{figure}

Figures \ref{fig:proband}(a) and (b) show the layer-projected band structures near the K point obtained by DFT calculations at $\theta=13.2^\circ$ and $\theta=5.1^\circ$, respectively. In the low-energy regime, \textit{aAB} features four bands composed of the Dirac bands originating from the twisted monolayer \textit{a} and two parabolic bands originating from the \textit{AB} bilayer. At these large twist angles, our calculations uncover a significant splitting $ \Delta $ of the parabolic bands of nearly 40 meV, which was not found by previous continuum model \cite{theory2} or tight-binding \cite{tight-binding} calculations. Moreover,  we find that $ \Delta $ decreases slightly with decreasing $\theta$ as shown in Fig. \ref{fig:proband}(c), which is different from the increasing trend previously reported \cite{tight-binding}. On the other hand, the Dirac point lies inside the gap $ \Delta $ of the parabolic bands at these angles.

From the perspective of the continuum model, the large momentum separation between the Dirac cone and the parabolic bands in the reciprocal space should result in a weak interaction between them. Therefore, significant splitting of the parabolic bands is not expected to occur at a large twist angle, especially in the absence of an external electric field. Here, we find this unexpected splitting of parabolic bands by spontaneous electric polarization (or an internal crystal field, equivalently) in the full-scale calculation. Although the \textit{AB} bilayer is weakly coupled with the twisted monolayer, the interlayer coupling can cause asymmetric charge transfer across layers \textit{A} and \textit{B} due to breaking of the inversion symmetry. From the layer-projected band structure in Figs.\;\ref{fig:proband}(a) and (b), it can be seen that the occupied state near the parabolic valence band maximum (VBM) is located mainly in layer \textit{A}, while the unoccupied state near the parabolic conduction band minimum (CBM) is located mainly in layer \textit{B}. This result implies a net charge transfer from layer \textit{B} to layer \textit{A}, creating a spontaneous electric polarization that points in the opposite direction, as shown in Fig.\;\ref{fig:proband}(d). The red and green $p_z$ orbitals show the layer- and sublattice-dependent states at the edge of the parabolic valence and conduction bands, respectively. Due to the direct alignment of sublattice $S_B$ in layer \textit{B} with sublattice $S_A$ in layer \textit{A}, their bonding and anti-bonding states do not appear in the low-energy regime. This visualization is valid for a large twist angle when the layer hybridization between \textit{a} and \textit{A} is weak. We confirm the presence of spontaneous polarization by directly calculating the vertical dipole moment through the Berry phase method \cite{King-Smith1993} in DFT calculations. For a large twist angle of $\theta=13.2^\circ$, we estimate the dipole moment to be $0.019\, e\cdot$\AA \; in one moiré supercell, corresponding to an electric polarization of 0.3 pC/m. 

At small twist angles, the \textit{aAB} moiré supercell becomes too large for DFT calculations. Therefore, we adopt an accurate TB model and compare the $\Delta$ value with the DFT results as shown in Fig.\;\ref{fig:proband}(c). The good consistency indicates that the TB model is accurate enough to capture the spontaneous electric polarization induced by the coupling between the twisted monolayer \textit{a} and the \textit{AB} bilayer.  We note that simplified TB parameters without angular dependence \cite{TB_parameter_1} would fail to capture the $\Delta$ value correctly, indicating the importance of an accurate description of the interlayer coupling at the twisted interface.

In calculations using the continuum model,  an internal crystal field should be added to the Hamiltonian in order to correctly capture the spontaneous polarization in the electronic structure. The physical origin of the split parabolic bands in \textit{aAB} is reminiscent of that in twisted double-bilayer graphene (TDBG), where an energy gap separates the conduction and valence parabolic bands with charge located in the outer layers and the inner layers, respectively \cite{TDBG1,TDBG2}. This phenomenon observed in TDBG was attributed to the internal crystal fields pointing from the surfaces towards the inner layers \cite{TDBG1,TDBG3}. In contrast to \textit{aAB}, however, the internal crystal fields in TDBG cancel out and do not generate a net electric polarization.

\subsection{Narrow bands near the quasi-magic angle}

\begin{figure*}[htb]
    \centering
    \includegraphics[width=0.92\textwidth]{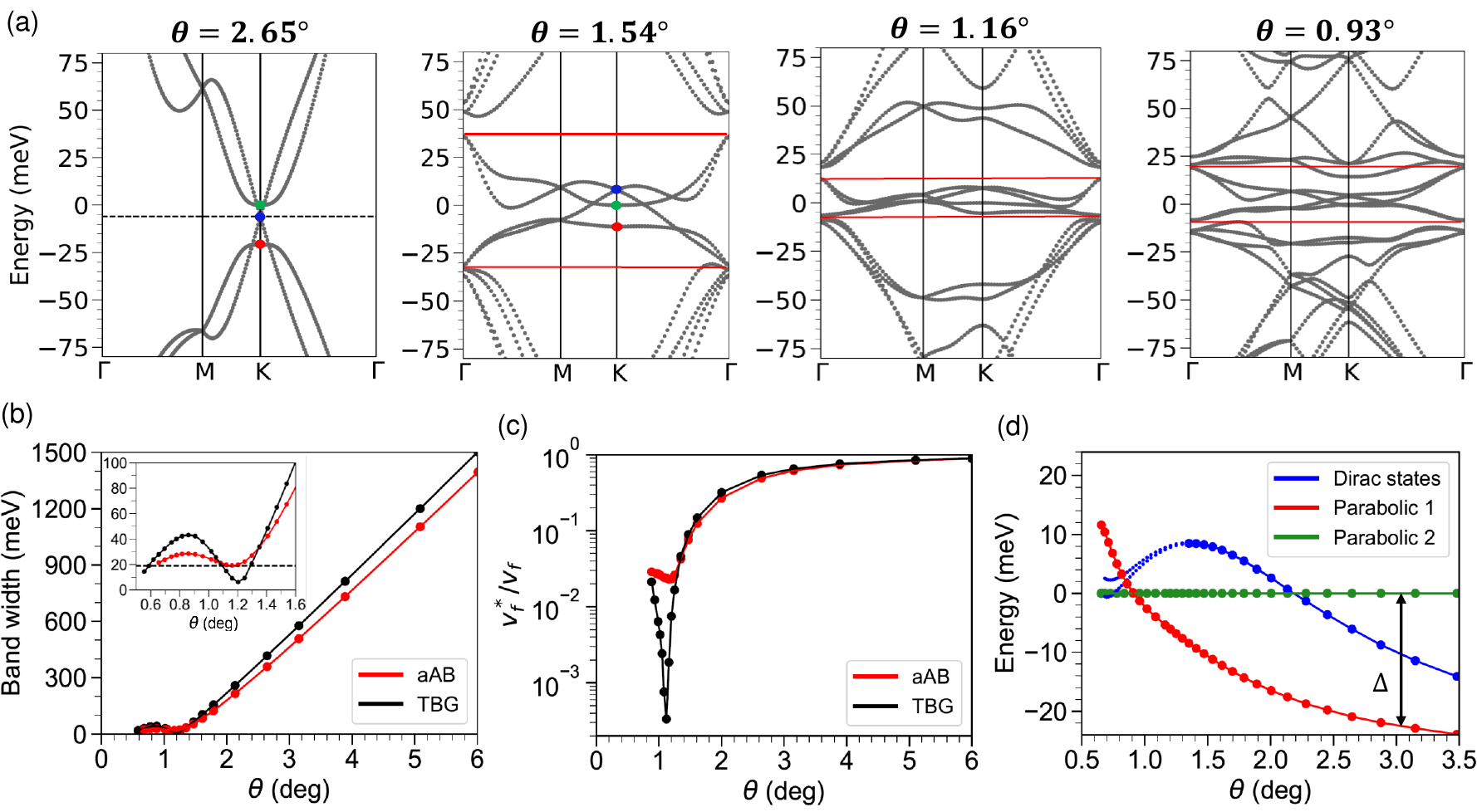}
    \caption{(a) Tight-binding band structures of \textit{aAB} at small twist angles, with the band width of the low-energy bands marked by red lines. The blue, red, and green dots label the Dirac states, parabolic state 1 and parabolic state 2, as in Fig.\;\ref{fig:proband}.  (b) Band width of the four low-energy bands of \textit{aAB} and TBG as a function of the twist angle. The inset shows that the \textit{aAB} low-energy bands reach a minimum width of about 19 meV (marked by a horizontal dashed line) at $\theta=1.16^\circ$, which we define as the quasi-magic angle (qMA). (c) Normalized Fermi velocity for the Dirac bands of \textit{aAB} and TBG at K. The Dirac bands of \textit{aAB} have a minimum slope at the qMA. (d) The energies of the four states at K as a function of the twist angle. The parabolic state 2 has a pinned energy, which is thus set to the zero reference. The splitting between the parabolic states is labeled as $\Delta$. The originally degenerate Dirac states split in energy at small angles, as represented by small blue dots.}
    \label{fig:bandstructure}
\end{figure*}

The band structures calculated using the accurate TB model for small $\theta$ angles between $0.93^\circ$ and $2.65^\circ$ are shown in Fig.\;\ref{fig:bandstructure}(a). As $\theta$ decreases, the parabolic bands and Dirac bands are considerably distorted, and their energy order also varies. Since these four low-energy bands are separated from the rest, a reasonable parameter to be considered is their whole band width. Figure \ref{fig:bandstructure}(b) shows that the band width does not decrease monotonically as a function of the twist angle. Instead, its variation highly resembles that in TBG. We find that the low-energy bands of \textit{aAB} reach a minimum width of approximately 19 meV at a twist angle of approximately $\theta = 1.16^\circ$, which is close to the first magic angle of the TBG. In addition, the normalized ``Fermi velocity'', calculated by the slope of the Dirac bands at the K point, also exhibits a minimum at the same twist angle, as shown in Fig.\;\ref{fig:bandstructure}(c). However, our calculations find that the normalized Fermi velocity does not vanish as that in TBG, and the four low-energy bands of \textit{aAB} are not as flat as those in TBG at the first magic angle, either. Therefore, we define this special angle as a quasi-magic angle (qMA) in \textit{aAB}. In the vicinity of the qMA, the band width of \textit{aAB} is slightly larger than that of TBG, but this situation reverses when the twist angle deviates about $0.1^\circ$ away from this value. 


Next, we analyze the energy variations of the four states at the K point when the twist angle decreases as shown in Fig.\;\ref{fig:bandstructure}(d). We change the notation of the parabolic valence and conduction bands at large angles to parabolic states 1 and 2, respectively, since they cross in energy at small twist angles. The positions of the Dirac point and parabolic states 1 and 2 in the band structure are marked by blue, red, and green dots, respectively, in Figs.\;\ref{fig:bandstructure}(a) and (d). Interestingly, our TB calculations reveal that the energy of parabolic state 2 remains pinned at a fixed value for all twist angles. Therefore, we set its energy to zero in Fig.\;\ref{fig:bandstructure}(a) and (d) as a reference point. This energy pinning arises from the characteristic wavefunction of parabolic state 2, which is confined solely to sublattice $S_A$ in layer \textit{B} [see our convention of sublattices in Fig.\;\ref{fig:proband}(d)]. 
For the $S_A$ site in bottom layer \textit{B}, its effective interlayer coupling to the adjacent layer \textit{A} nearly cancels out at the K point, and its interaction with the twisted top monolayer \textit{a} is neglected in the TB model. Consequently, the parabolic state 2 at K is completely decoupled from the other states, and thus its wavefunction remains unchanged with a pinned eigenvalue for all twist angles. On the other hand, the Dirac states from monolayer \textit{a} shift upward in energy as $\theta$ decreases \cite{theory1,theory2, tight-binding} and split around $\theta \approx 1.3^\circ$, where the Dirac cone reaches the highest energy position.

While parabolic state 2 at K is completely distributed in layer \textit{B} regardless of the twist angle,  at small twist angles parabolic state 1 and the Dirac state are no longer confined in the corresponding layers of  \textit{A} and \textit{a},  due to increased layer hybridization. The layer-projected band structure calculated using the TB model is shown in Fig. 6 in Appendix A. For $\theta$ near the qMA, the four narrow bands can hardly be identified individually as parabolic bands corresponding to the \textit{AB} bilayer and Dirac bands corresponding to the \textit{a} monolayer. As will be further discussed in the next section, at small twist angles the \textit{aAB} trilayer should be regarded as TBG interacting with layer \textit{B}, due to the strong enhancement of interlayer interaction at the twisted interface.

\subsection{Layer-dependent charge localization and delocalization}

\begin{figure}[htb]
    \centering
    \includegraphics[width=0.48\textwidth]{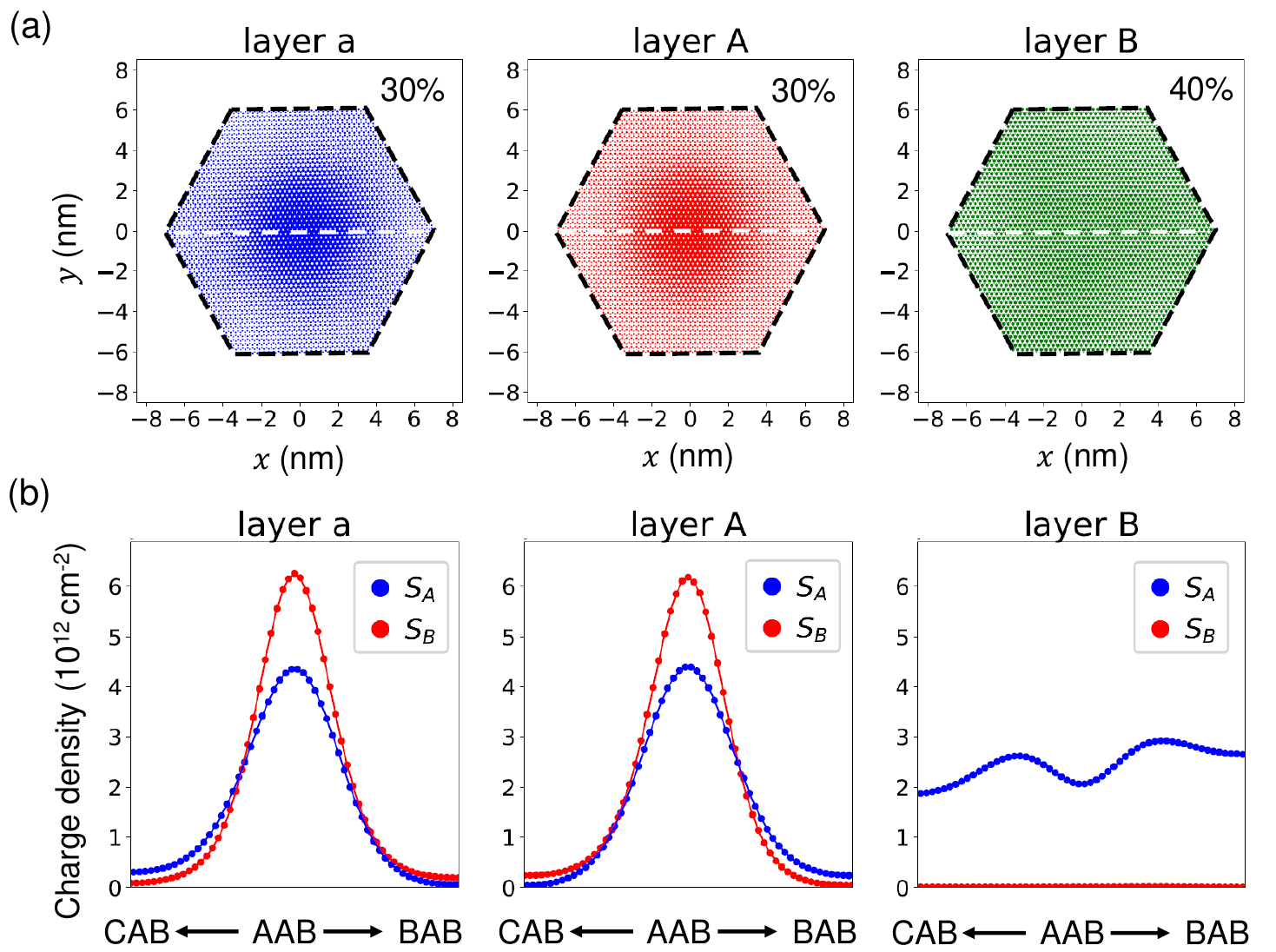}
    \caption{(a) Layer-resolved charge distribution integrated over the four narrow bands of \textit{aAB} at $\theta=1.16^{\circ}$ calculated by the tight-binding method. The charge is localized in the \textit{AAB} region in \textit{a} and \textit{A}  layers while delocalized in layer B. (b) Sublattice-resolved charge distribution (per spin) along the white dashed line in (a), going through the \textit{CAB}, \textit{AAB} to \textit{BAB} regions (see Fig.\;\ref{fig:moire}). The charge distribution in layer \textit{B} is completely sublattice polarized.}
    \label{fig:charge}
\end{figure}

The observed charge localization in the AA region of TBG significantly enhances the Coulomb interaction between electrons \cite{TBG_localization_1,TBG_localization_2,TBG_localization_3}. Here we study the narrow bands of \textit{aAB} near the Fermi level by examining the layer-resolved charge distribution at the qMA  ($\theta=1.16^\circ$) as shown in Fig.\;\ref{fig:charge}(a). By integrating over the Brillouin zone and summing over the four low-energy bands, we find that the adjacent twisted layers (\textit{a} and \textit{A} ) exhibit clear charge localization in the \textit{AAB} region, consistent with the findings by scanning tunneling microscopy \cite{Ling_2022, Si_2022}. In contrast, the charge distribution in the outer bilayer (layer \textit{B}) is delocalized. Our results verify that charge localization and delocalization coexist within the four low-energy bands of the \textit{aAB} system \cite{Ling_2022}.  At the qMA, approximately 40\% of the charge is distributed in layer B, and 30\% in each of the other layers. 

Figure\;\ref{fig:charge}(b) further illustrates the sublattice-dependent charge distribution along the high-symmetry line in real space from \textit{CAB}, \textit{AAB} to \textit{BAB} regions as shown in Fig.\;\ref{fig:moire}. We find that the charge distribution in layer \textit{B} is completely sublattice polarized on sublattice $S_A$. This special charge distribution implies the non-trivial topology of \textit{aAB} narrow bands. Moreover, the charge localization patterns in \textit{a}  and \textit{A} layers are nearly identical, forming TBG-like bilayers. Our results suggest that the interlayer coupling at the twisted interface generates a TBG-like moiré potential that traps low-energy electrons, and this moiré potential becomes dominant as  $\theta$ approaches the qMA. Consequently, instead of considering the \textit{aAB} trilayer as the \textit{AB} bilayer interacting with a twisted monolayer \textit{a}, it should be regarded as TBG interacting with a single layer \textit{B}. The existence of layer \textit{B} breaks the sublattice symmetry in the TBG layers through a sublattice-dependent potential. Due to the alignment of the sublattices as shown in Fig.\;\ref{fig:proband}(d), the charge in layer \textit{B} vanishes on $S_B$ sites in the low-energy regime. On the other hand, the TBG-like layers should be considered as a single system with charge distribution on both $S_A$ and $S_B$ sublattices but a finite sublattice polarization due to the influence of layer \textit{B}.

\subsection{Evolution of interlayer couplings}
\begin{figure*}[htb]
    \centering
    \includegraphics[width=0.92\textwidth]{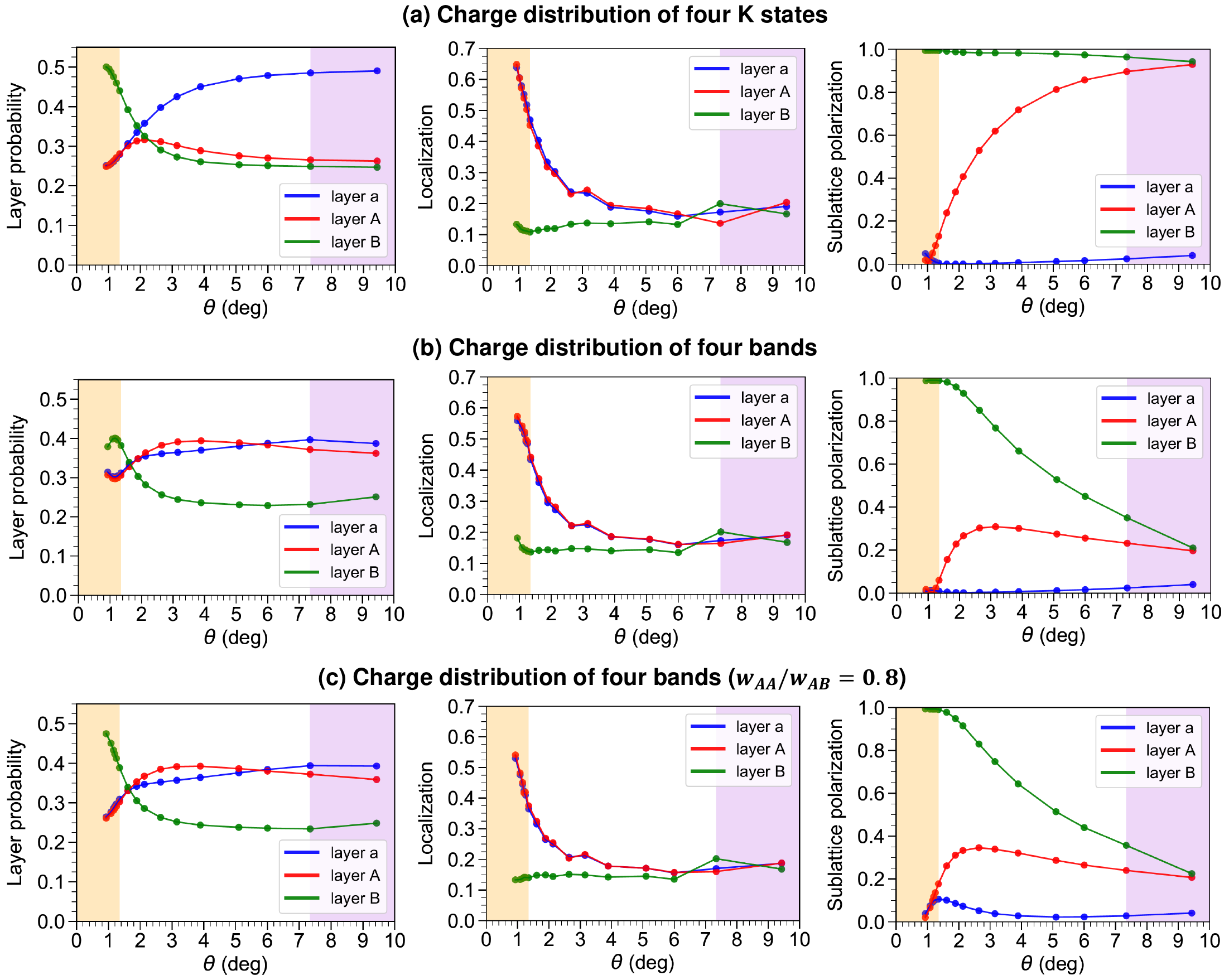}
    \caption{Evolution of layer-dependent charge distribution in \textit{aAB} as a function of the twist angle. We consider three quantities in these figures: layer probability, charge localization (see text), and sublattice polarization $p$. The results are presented by summing over (a) the four low-energy states at the K point, (b) the four low-energy bands, and (c) the four low-energy bands with $w_{AA}/w_{AB}=0.8$. The purple and yellow shaded regions mark the large ($\theta >7.3^\circ$) and small ($\theta <1.3^\circ$) twist-angle regimes, respectively.}
    \label{fig:Evolution}
\end{figure*}
In trilayer systems with two interfaces, whenever one interlayer coupling dominates the other, the corresponding bilayer should be considered approximately as one system, weakly interacting with the additional monolayer. Although we have demonstrated the two distinct behaviors of the low-energy bands in the \textit{aAB} trilayer at large twist angles and near the qMA in previous sections, we will further illustrate the competition and evolution of interlayer interactions in this section by analyzing the layer-resolved charge distribution at different twist angles obtained by tight-binding calculations. Three different features will be presented to illustrate the quantitative evolution of the charge distribution. First, the layer probability distribution represents the charge occupation in each layer. Second, a charge localization measure is defined by the charge fraction in the \textit{AAB} region within a radius $r<0.2L$, where $L$ is the moiré lattice constant. Third, sublattice polarization is defined as $p_i=|S_{A,i}-S_{B,i}|/(S_{A,i}+S_{B,i})$, where $S_{\sigma,i}$ is the total charge on sublattice $\sigma$ in layer $i$.  
 
Since the behavior of low-energy electrons is of primary interest, we first examine the charge distribution summed over the four states at K. Figure \ref{fig:Evolution}(a) shows that in the large twist-angle regime (the purple region with $\theta>7.3^\circ$), \textit{aAB} behaves as an \textit{AB} bilayer weakly coupled with the twisted monolayer \textit{a}, resulting in a sublattice-polarized \textit{AB} bilayer with  $p\approx 1$ and sublattice-unpolarized monolayer \textit{a} with $p\approx 0$. In this angle range, the layer probability is approximately 2:1:1 for the \textit{a}, \textit{A}, and \textit{B} layers, respectively, corresponding to the degenerate Dirac states in monolayer \textit{a} and the split parabolic states in the \textit{AB} bilayer. The charge localization behavior in each layer approaches the uniform distribution limit, which is 0.145 from our definition.  In contrast, in the small-angle regime (the yellow region with $\theta<1.3^\circ$), the \textit{a} and \textit{A}  layers are strongly coupled, exhibiting the same layer probability, large localization value, and vanishing sublattice polarization $p$, as shown by the blue and red curves in Fig.\;\ref{fig:Evolution}(a) . These features indicate the formation of TBG-like layers, with the layer charge ratio becoming roughly 1:1:2 for the \textit{a}, \textit{A}, and \textit{B} layers, respectively. The two twist-angle regimes marked with yellow and purple in Fig.\;\ref{fig:Evolution} demonstrate distinct properties of the \textit{aAB} trilayer.   

In the intermediate twist-angle regime, the sublattice polarization of layer \textit{a}  and layer \textit{B} remains at $p=0$ and $p=1$, respectively.  For the middle layer \textit{A}, its sublattice polarization decreases with decreasing $\theta$  due to the enhanced interlayer coupling at the twisted interface, reflecting the competition of interlayer interactions. We estimate that layer A's sublattice polarization decreases to $ 1/2$ at a critical angle of about $\theta^*=2.5^\circ$, which can be identified as a crossover of the interlayer couplings. In addition, while the charge localization increases tremendously when $\theta<\theta^*$, the localization values of \textit{a} and \textit{A} layers coincide much earlier and stay the same throughout the intermediate twist angle regime, suggesting that a similar two-dimensional (2D) moiré potential has already been present in these two layers,  even when their layer probability and $p$ differ greatly.  However, this moiré potential is not strong enough to completely couple layers \textit{a} and \textit{A}, leaving a finite sublattice polarization in layer A until $\theta$ approaches the small-angle regime. The real-space charge distribution patterns at various twist angles are presented in Fig. 7 in Appendix B. 

We also calculated the total charge distribution of the four low-energy bands. The results are shown in Fig.\;\ref{fig:Evolution}(b). In the large-angle regime, the layer probability significantly differs from that calculated at the K point due to the wider energy window. For the same reason, the sublattice polarizations of layers \textit{A} and \textit{B} deviate from 1, but still coincide. As the energy window narrows with a decreasing twist angle,  the sublattice polarization of layer \textit{B} reasonably increases to 1, while the sublattice polarization of layer \textit{A} is influence in an opposite way by the interlayer couplings at the two interfaces: the coupling to layer \textit{B} increases $p$, while the coupling to layer \textit{a} reduces it. A turning point is around $\theta^*=2.5^\circ$, below which the interlayer coupling effect at the twisted interfaces is enhanced significantly, leading to a drastic decrease of layer \textit{A}'s sublattice polarization to zero. In the end, layers \textit{a} and \textit{A} become strongly coupled in the small twist angle regime, resembling the results obtained at the K-point alone. 

So far we have considered a trilayer system in which the interlayer separation is uniform. However, corrugation can occur between the twisted pair of layers \textit{a} and \textit{A},  with a larger layer separation in the AA-stacked region compared to that in the AB-stacked region. To simulate this lattice relaxation effect, one can reduce the intra-sublattice coupling $w_{AA}$ with respect to the inter-sublattice coupling $w_{AB}$ at the twisted interface. We chose a ratio of $w_{AA}/w_{AB}=0.8$  in our TB calculations, and 
resulting charge distributions are shown in Fig.\;\ref{fig:Evolution}(c). We find that the charge density and localization value in each layer do not change much as compared with results in Fig.\;\ref{fig:Evolution}(b) with $w_{AA}/w_{AB}=1$, except that the sublattice polarization of layer \textit{a} is slightly increased below $\theta=3^\circ$ 
before coinciding with that of layer \textit{A} and decreasing together at small $\theta$ values. 


Our results in Fig.\;\ref{fig:Evolution} provide evidence for the evolution of interlayer couplings in the \textit{aAB} trilayer system. As the twist angle decreases, the interaction between the AB bilayer remains fixed, but the interlayer interaction effect at the twisted interface increases dramatically. At a crossover angle around $\theta^*=2.5^\circ$, the sublattice polarization $p$ of layer \textit{A} drops to $p=1/2$ for the four K-point states,  and $p$ also reaches a turning point at $\theta^*$ for the total charge distribution within the four bands. This crossover represents a balance between the two interlayer interactions, showing equal influence over the charge distribution at the atomic scale on the middle layer. Below $\theta^*$, layer \textit{A}'s sublattice polarization reduces significantly, and the charge localization is greatly enhanced due to the strong moiré potential at the twisted interface. When the twist angle is sufficiently small ($\theta <1.3^\circ$), the moiré potential dominates, leading to TBG-like layers with $p\approx 0$ and monolayer \textit{B} with $p\approx 1$. 

\section{Conclusions}
In this work, we employ first-principles calculations combined with an accurate tight-binding model to investigate the electronic properties of the \textit{aAB} trilayer beyond the continuum model, focusing on the evolution and competition of interlayer interaction effects at two interfaces. Our study reveals two distinct electronic regimes depending on the twist angle $\theta$. At large $\theta$, the \textit{aAB} trilayer behaves like a Bernal \textit{AB} bilayer with parabolic bands interacting with a twisted monolayer \textit{a} with linear Dirac bands. Interlayer interactions induce a spontaneous electric polarization within the \textit{AB} bilayer, splitting the parabolic bands by approximately 30–40 meV, which can be accurately described using tight-binding parameters with angular dependence. As $\theta$ decreases, the interlayer coupling effect between layers \textit{a} and \textit{A} strengthens, generating a 2D moiré potential that traps electrons and reduces the sublattice polarization $p$ of the middle layer \textit{A}. In the small twist-angle regime ($\theta <1.3^\circ$), where correlated physics emerges in experiments, layers \textit{a} and \textit{A} are more strongly coupled and resemble twisted bilayer graphene with charge localization in the \textit{AAB} regions and $p\approx0$. In contrast, layer \textit{B} exhibits a delocalized charge distribution with complete sublattice polarization ($p=1$). We conclude that at sufficiently small $\theta$, the \textit{aAB} trilayer is better described as twisted bilayer graphene interacting with monolayer \textit{B}. Namely, \textit{aAB} retains the key electronic features of TBG, including having a minimum width of the narrow bands and a minimum Fermi velocity of the Dirac bands at a quasi-magic angle of $\theta =1.16^\circ$, closely matching the first magic angle of twisted bilayer graphene.

\begin{acknowledgments}
The authors thank Martin Callsen and Chi-Ruei Pan for helpful discussions. This work is supported by Academia Sinica, Taiwan.
\end{acknowledgments}

\appendix

\section{Layer-projected band structure}
\begin{figure*}[htbp]
    \centering
    \includegraphics[width=0.9\textwidth]{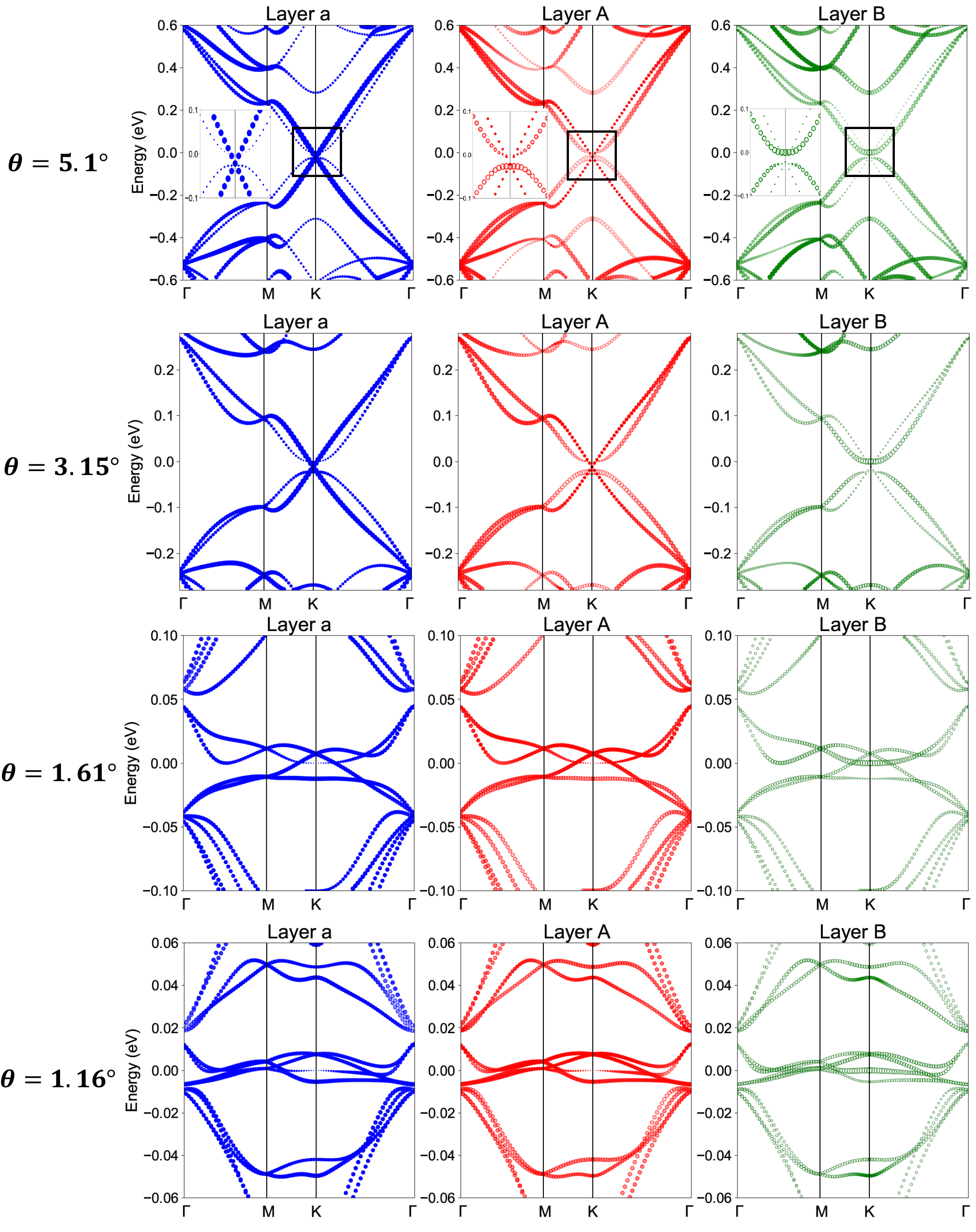}
    \caption{Layer-projected band structures of \textit{aAB} at various twist angles, calculated using the tight-binding model. Each state in the band structure is represented by a circle, where the circle size indicates the layer probability, and the transparency of the circle represents sublattice polarization. Fully transparent circles correspond to states with sublattice polarization $p\approx 1$, and vice versa. }
    \label{fig:proband_TB}
\end{figure*}

Figure \ref{fig:proband_TB} shows the layer-projected band structures of \textit{aAB} for various twist angles calculated using the tight-binding model. Each state in the band structure is represented by a circle, where the circle size indicates the layer probability, and the transparency of the circle represents sublattice polarization. Sublattice polarization is defined as $p_i=|S_{A,i}-S_{B,i}|/(S_{A,i}+S_{B,i})$, where $S_{\sigma,i}$ is the total charge on sublattice $\sigma$ in layer \textit{i}. Fully transparent circles correspond to states with sublattice polarization $p\approx 1$, and vice versa. 

The inset for $\theta=5.1^\circ$ shows the projected low-energy bands near the K point. At this twist angle, the Dirac bands are primarily contributed by layer \textit{a} with $p=0$, and the parabolic VBM and CBM states are predominantly from layers \textit{A} and \textit{B}, respectively, with $p=1$. As the twist angle decreases, the Dirac states and the parabolic VBM state have increasing contributions from the other layers due to the enhancement of layer hybridization. Notably, the parabolic CBM state (set as energy zero) is still localized only on layer \textit{B} for all twist angles, since it is decoupled from the other states in the TB model, as explained in the main text. 

The evolution of sublattice polarization for the four bands can also be observed from Fig. \ref{fig:proband_TB}. When projected onto layer \textit{a}, the four bands remain sublattice-unpolarized with $p\approx0$ for all twist angles. For layer \textit{B} at a large twist angle, the four bands are nearly completely sublattice polarized ($p=1$) in the low-energy regime with $|E|<0.1$ eV, and $p$ deviates from 1 in a wider energy window. As $\theta$ decreases, layer \textit{B}'s sublattice polarization for four bands increases, approaching $p=1$ due to the narrowing of the energy window. On the other hand, at $\theta=5.1^\circ$,  layer \textit{A} shows significant sublattice polarization in the low-energy regime $|E|<0.1$ eV, similar to the situation in layer \textit{B}. However, at a small twist angle of $\theta=1.16^\circ$, layer \textit{A}'s sublattice polarization decreases and becomes similar to that of layer \textit{a}.

\section{Evolution of charge distribution}

\begin{figure*}[htbp]
    \centering
    \includegraphics[width=0.85\textwidth]{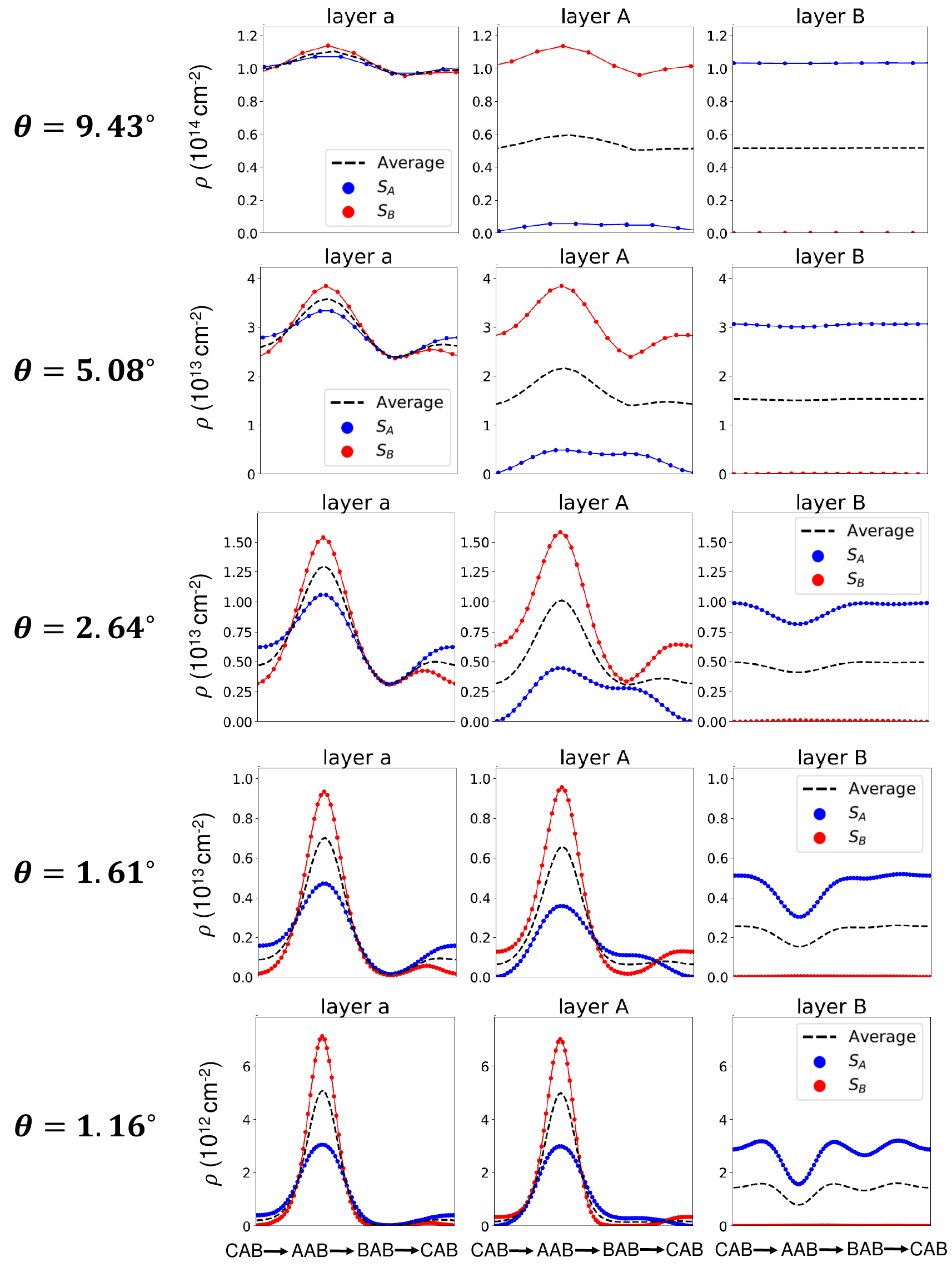}
    \caption{Evolution of layer-resolved charge distribution summed over the four states at K as a function of the twist angle. The charge density $\rho$ per spin is calculated by the tight-binding method along the high-symmetry line, going through the \textit{CAB}, \textit{AAB}, \textit{BAB} to \textit{CAB} regions. The charge distribution on sublattice A and B is marked by blue and red, respectively, while their average is shown by black dashed lines.}
    \label{fig:K_charge}
\end{figure*}

The evolution of the real-space charge distribution for the states at K is shown in Fig.\;\ref{fig:K_charge}. At a large twist angle $\theta=9.43^\circ$, the charge density in each layer is nearly uniform, with a layer probability ratio of approximately 2:1:1 for layers \textit{a}, \textit{A}, and \textit{B}. At the atomic scale, the charge distribution in layers \textit{A} and \textit{B} is sublattice polarized at different sublattices, while layer \textit{a} is not sublattice polarized. The result indicates that the \textit{AB} bilayer should be considered as one entity that weakly couples with the twisted monolayer \textit{a}. 

As the twist angle decreases, the 2D moiré potential is enhanced and charge localization  develops in the \textit{AAB} region of layers \textit{a} and \textit{A}. Notably, the average charge distribution patterns in layers \textit{a} and \textit{A} are highly similar, suggesting that they are influenced by the same 2D moiré potential.  In the intermediate twist-angle regime, however, the moiré potential is not strong enough to completely couple layers \textit{a} and \textit{A}, leaving finite sublattice polarization in layer \textit{A}. At sufficiently small twist angles, the interlayer coupling effect at the twisted interface is strengthened and the charge distribution in layers \textit{a} and \textit{A} are nearly identical even at the atomic scale, forming TBG-like layers. On the other hand, the charge distribution in layer \textit{B} remains completely sublattice polarized. Our results show that \textit{aAB} should be considered as TBG interacting with layer \textit{B} at small twist angles.

\clearpage

\bibliography{aAB}

\begin{thebibliography}{40}%
\makeatletter
\providecommand \@ifxundefined [1]{%
 \@ifx{#1\undefined}
}%
\providecommand \@ifnum [1]{%
 \ifnum #1\expandafter \@firstoftwo
 \else \expandafter \@secondoftwo
 \fi
}%
\providecommand \@ifx [1]{%
 \ifx #1\expandafter \@firstoftwo
 \else \expandafter \@secondoftwo
 \fi
}%
\providecommand \natexlab [1]{#1}%
\providecommand \enquote  [1]{``#1''}%
\providecommand \bibnamefont  [1]{#1}%
\providecommand \bibfnamefont [1]{#1}%
\providecommand \citenamefont [1]{#1}%
\providecommand \href@noop [0]{\@secondoftwo}%
\providecommand \href [0]{\begingroup \@sanitize@url \@href}%
\providecommand \@href[1]{\@@startlink{#1}\@@href}%
\providecommand \@@href[1]{\endgroup#1\@@endlink}%
\providecommand \@sanitize@url [0]{\catcode `\\12\catcode `\$12\catcode `\&12\catcode `\#12\catcode `\^12\catcode `\_12\catcode `\%12\relax}%
\providecommand \@@startlink[1]{}%
\providecommand \@@endlink[0]{}%
\providecommand \url  [0]{\begingroup\@sanitize@url \@url }%
\providecommand \@url [1]{\endgroup\@href {#1}{\urlprefix }}%
\providecommand \urlprefix  [0]{URL }%
\providecommand \Eprint [0]{\href }%
\providecommand \doibase [0]{https://doi.org/}%
\providecommand \selectlanguage [0]{\@gobble}%
\providecommand \bibinfo  [0]{\@secondoftwo}%
\providecommand \bibfield  [0]{\@secondoftwo}%
\providecommand \translation [1]{[#1]}%
\providecommand \BibitemOpen [0]{}%
\providecommand \bibitemStop [0]{}%
\providecommand \bibitemNoStop [0]{.\EOS\space}%
\providecommand \EOS [0]{\spacefactor3000\relax}%
\providecommand \BibitemShut  [1]{\csname bibitem#1\endcsname}%
\let\auto@bib@innerbib\@empty
\bibitem [{\citenamefont {Cao}\ \emph {et~al.}(2018{\natexlab{a}})\citenamefont {Cao}, \citenamefont {Fatemi}, \citenamefont {Demir}, \citenamefont {Fang}, \citenamefont {Tomarken}, \citenamefont {Luo}, \citenamefont {Sanchez-Yamagishi}, \citenamefont {Watanabe}, \citenamefont {Taniguchi}, \citenamefont {Kaxiras}, \citenamefont {Ashoori},\ and\ \citenamefont {Jarillo-Herrero}}]{correlated}%
  \BibitemOpen
  \bibfield  {author} {\bibinfo {author} {\bibfnamefont {Y.}~\bibnamefont {Cao}}, \bibinfo {author} {\bibfnamefont {V.}~\bibnamefont {Fatemi}}, \bibinfo {author} {\bibfnamefont {A.}~\bibnamefont {Demir}}, \bibinfo {author} {\bibfnamefont {S.}~\bibnamefont {Fang}}, \bibinfo {author} {\bibfnamefont {S.~L.}\ \bibnamefont {Tomarken}}, \bibinfo {author} {\bibfnamefont {J.~Y.}\ \bibnamefont {Luo}}, \bibinfo {author} {\bibfnamefont {J.~D.}\ \bibnamefont {Sanchez-Yamagishi}}, \bibinfo {author} {\bibfnamefont {K.}~\bibnamefont {Watanabe}}, \bibinfo {author} {\bibfnamefont {T.}~\bibnamefont {Taniguchi}}, \bibinfo {author} {\bibfnamefont {E.}~\bibnamefont {Kaxiras}}, \bibinfo {author} {\bibfnamefont {R.~C.}\ \bibnamefont {Ashoori}},\ and\ \bibinfo {author} {\bibfnamefont {P.}~\bibnamefont {Jarillo-Herrero}},\ }\bibfield  {title} {\bibinfo {title} {Correlated insulator behaviour at half-filling in magic-angle graphene superlattices},\ }\href {https://doi.org/10.1038/nature26154} {\bibfield  {journal} {\bibinfo  {journal}
  {Nature}\ }\textbf {\bibinfo {volume} {556}},\ \bibinfo {pages} {80–84} (\bibinfo {year} {2018}{\natexlab{a}})}\BibitemShut {NoStop}%
\bibitem [{\citenamefont {Lu}\ \emph {et~al.}(2019)\citenamefont {Lu}, \citenamefont {Stepanov}, \citenamefont {Yang}, \citenamefont {Xie}, \citenamefont {Aamir}, \citenamefont {Das}, \citenamefont {Urgell}, \citenamefont {Watanabe}, \citenamefont {Taniguchi}, \citenamefont {Zhang}, \citenamefont {Bachtold}, \citenamefont {MacDonald},\ and\ \citenamefont {Efetov}}]{correlated+sc+mag2}%
  \BibitemOpen
  \bibfield  {author} {\bibinfo {author} {\bibfnamefont {X.}~\bibnamefont {Lu}}, \bibinfo {author} {\bibfnamefont {P.}~\bibnamefont {Stepanov}}, \bibinfo {author} {\bibfnamefont {W.}~\bibnamefont {Yang}}, \bibinfo {author} {\bibfnamefont {M.}~\bibnamefont {Xie}}, \bibinfo {author} {\bibfnamefont {M.~A.}\ \bibnamefont {Aamir}}, \bibinfo {author} {\bibfnamefont {I.}~\bibnamefont {Das}}, \bibinfo {author} {\bibfnamefont {C.}~\bibnamefont {Urgell}}, \bibinfo {author} {\bibfnamefont {K.}~\bibnamefont {Watanabe}}, \bibinfo {author} {\bibfnamefont {T.}~\bibnamefont {Taniguchi}}, \bibinfo {author} {\bibfnamefont {G.}~\bibnamefont {Zhang}}, \bibinfo {author} {\bibfnamefont {A.}~\bibnamefont {Bachtold}}, \bibinfo {author} {\bibfnamefont {A.~H.}\ \bibnamefont {MacDonald}},\ and\ \bibinfo {author} {\bibfnamefont {D.~K.}\ \bibnamefont {Efetov}},\ }\bibfield  {title} {\bibinfo {title} {Superconductors, orbital magnets and correlated states in magic-angle bilayer graphene},\ }\href
  {https://doi.org/10.1038/s41586-019-1695-0} {\bibfield  {journal} {\bibinfo  {journal} {Nature}\ }\textbf {\bibinfo {volume} {574}},\ \bibinfo {pages} {653–657} (\bibinfo {year} {2019})}\BibitemShut {NoStop}%
\bibitem [{\citenamefont {Cao}\ \emph {et~al.}(2018{\natexlab{b}})\citenamefont {Cao}, \citenamefont {Fatemi}, \citenamefont {Fang}, \citenamefont {Watanabe}, \citenamefont {Taniguchi}, \citenamefont {Kaxiras},\ and\ \citenamefont {Jarillo-Herrero}}]{sc+TBG}%
  \BibitemOpen
  \bibfield  {author} {\bibinfo {author} {\bibfnamefont {Y.}~\bibnamefont {Cao}}, \bibinfo {author} {\bibfnamefont {V.}~\bibnamefont {Fatemi}}, \bibinfo {author} {\bibfnamefont {S.}~\bibnamefont {Fang}}, \bibinfo {author} {\bibfnamefont {K.}~\bibnamefont {Watanabe}}, \bibinfo {author} {\bibfnamefont {T.}~\bibnamefont {Taniguchi}}, \bibinfo {author} {\bibfnamefont {E.}~\bibnamefont {Kaxiras}},\ and\ \bibinfo {author} {\bibfnamefont {P.}~\bibnamefont {Jarillo-Herrero}},\ }\bibfield  {title} {\bibinfo {title} {Unconventional superconductivity in magic-angle graphene superlattices},\ }\href {https://doi.org/10.1038/nature26160} {\bibfield  {journal} {\bibinfo  {journal} {Nature}\ }\textbf {\bibinfo {volume} {556}},\ \bibinfo {pages} {43–50} (\bibinfo {year} {2018}{\natexlab{b}})}\BibitemShut {NoStop}%
\bibitem [{\citenamefont {Sharpe}\ \emph {et~al.}(2019)\citenamefont {Sharpe}, \citenamefont {Fox}, \citenamefont {Barnard}, \citenamefont {Finney}, \citenamefont {Watanabe}, \citenamefont {Taniguchi}, \citenamefont {Kastner},\ and\ \citenamefont {Goldhaber-Gordon}}]{mag1}%
  \BibitemOpen
  \bibfield  {author} {\bibinfo {author} {\bibfnamefont {A.~L.}\ \bibnamefont {Sharpe}}, \bibinfo {author} {\bibfnamefont {E.~J.}\ \bibnamefont {Fox}}, \bibinfo {author} {\bibfnamefont {A.~W.}\ \bibnamefont {Barnard}}, \bibinfo {author} {\bibfnamefont {J.}~\bibnamefont {Finney}}, \bibinfo {author} {\bibfnamefont {K.}~\bibnamefont {Watanabe}}, \bibinfo {author} {\bibfnamefont {T.}~\bibnamefont {Taniguchi}}, \bibinfo {author} {\bibfnamefont {M.~A.}\ \bibnamefont {Kastner}},\ and\ \bibinfo {author} {\bibfnamefont {D.}~\bibnamefont {Goldhaber-Gordon}},\ }\bibfield  {title} {\bibinfo {title} {Emergent ferromagnetism near three-quarters filling in twisted bilayer graphene},\ }\href {https://doi.org/10.1126/science.aaw3780} {\bibfield  {journal} {\bibinfo  {journal} {Science}\ }\textbf {\bibinfo {volume} {365}},\ \bibinfo {pages} {605} (\bibinfo {year} {2019})}\BibitemShut {NoStop}%
\bibitem [{\citenamefont {Nuckolls}\ \emph {et~al.}(2020)\citenamefont {Nuckolls}, \citenamefont {Oh}, \citenamefont {Wong}, \citenamefont {Lian}, \citenamefont {Watanabe}, \citenamefont {Taniguchi}, \citenamefont {Bernevig},\ and\ \citenamefont {Yazdani}}]{Chern1}%
  \BibitemOpen
  \bibfield  {author} {\bibinfo {author} {\bibfnamefont {K.~P.}\ \bibnamefont {Nuckolls}}, \bibinfo {author} {\bibfnamefont {M.}~\bibnamefont {Oh}}, \bibinfo {author} {\bibfnamefont {D.}~\bibnamefont {Wong}}, \bibinfo {author} {\bibfnamefont {B.}~\bibnamefont {Lian}}, \bibinfo {author} {\bibfnamefont {K.}~\bibnamefont {Watanabe}}, \bibinfo {author} {\bibfnamefont {T.}~\bibnamefont {Taniguchi}}, \bibinfo {author} {\bibfnamefont {B.~A.}\ \bibnamefont {Bernevig}},\ and\ \bibinfo {author} {\bibfnamefont {A.}~\bibnamefont {Yazdani}},\ }\bibfield  {title} {\bibinfo {title} {Strongly correlated chern insulators in magic-angle twisted bilayer graphene},\ }\href {https://doi.org/10.1038/s41586-020-3028-8} {\bibfield  {journal} {\bibinfo  {journal} {Nature}\ }\textbf {\bibinfo {volume} {588}},\ \bibinfo {pages} {610–615} (\bibinfo {year} {2020})}\BibitemShut {NoStop}%
\bibitem [{\citenamefont {Wu}\ \emph {et~al.}(2021)\citenamefont {Wu}, \citenamefont {Zhang}, \citenamefont {Watanabe}, \citenamefont {Taniguchi},\ and\ \citenamefont {Andrei}}]{Chern2}%
  \BibitemOpen
  \bibfield  {author} {\bibinfo {author} {\bibfnamefont {S.}~\bibnamefont {Wu}}, \bibinfo {author} {\bibfnamefont {Z.}~\bibnamefont {Zhang}}, \bibinfo {author} {\bibfnamefont {K.}~\bibnamefont {Watanabe}}, \bibinfo {author} {\bibfnamefont {T.}~\bibnamefont {Taniguchi}},\ and\ \bibinfo {author} {\bibfnamefont {E.~Y.}\ \bibnamefont {Andrei}},\ }\bibfield  {title} {\bibinfo {title} {Chern insulators, van hove singularities and topological flat bands in magic-angle twisted bilayer graphene},\ }\href {https://doi.org/10.1038/s41563-020-00911-2} {\bibfield  {journal} {\bibinfo  {journal} {Nature Materials}\ }\textbf {\bibinfo {volume} {20}},\ \bibinfo {pages} {488–494} (\bibinfo {year} {2021})}\BibitemShut {NoStop}%
\bibitem [{\citenamefont {Xie}\ \emph {et~al.}(2021)\citenamefont {Xie}, \citenamefont {Pierce}, \citenamefont {Park}, \citenamefont {Parker}, \citenamefont {Khalaf}, \citenamefont {Ledwith}, \citenamefont {Cao}, \citenamefont {Lee}, \citenamefont {Chen}, \citenamefont {Forrester} \emph {et~al.}}]{Chern3}%
  \BibitemOpen
  \bibfield  {author} {\bibinfo {author} {\bibfnamefont {Y.}~\bibnamefont {Xie}}, \bibinfo {author} {\bibfnamefont {A.~T.}\ \bibnamefont {Pierce}}, \bibinfo {author} {\bibfnamefont {J.~M.}\ \bibnamefont {Park}}, \bibinfo {author} {\bibfnamefont {D.~E.}\ \bibnamefont {Parker}}, \bibinfo {author} {\bibfnamefont {E.}~\bibnamefont {Khalaf}}, \bibinfo {author} {\bibfnamefont {P.}~\bibnamefont {Ledwith}}, \bibinfo {author} {\bibfnamefont {Y.}~\bibnamefont {Cao}}, \bibinfo {author} {\bibfnamefont {S.~H.}\ \bibnamefont {Lee}}, \bibinfo {author} {\bibfnamefont {S.}~\bibnamefont {Chen}}, \bibinfo {author} {\bibfnamefont {P.~R.}\ \bibnamefont {Forrester}}, \emph {et~al.},\ }\bibfield  {title} {\bibinfo {title} {Fractional chern insulators in magic-angle twisted bilayer graphene},\ }\href {https://doi.org/10.1038/s41586-021-04002-3} {\bibfield  {journal} {\bibinfo  {journal} {Nature}\ }\textbf {\bibinfo {volume} {600}},\ \bibinfo {pages} {439–443} (\bibinfo {year} {2021})}\BibitemShut {NoStop}%
\bibitem [{\citenamefont {Bistritzer}\ and\ \citenamefont {MacDonald}(2011)}]{Bistritzer2011}%
  \BibitemOpen
  \bibfield  {author} {\bibinfo {author} {\bibfnamefont {R.}~\bibnamefont {Bistritzer}}\ and\ \bibinfo {author} {\bibfnamefont {A.~H.}\ \bibnamefont {MacDonald}},\ }\bibfield  {title} {\bibinfo {title} {Moiré bands in twisted double-layer graphene},\ }\href {https://doi.org/10.1073/pnas.1108174108} {\bibfield  {journal} {\bibinfo  {journal} {Proceedings of the National Academy of Sciences}\ }\textbf {\bibinfo {volume} {108}},\ \bibinfo {pages} {12233} (\bibinfo {year} {2011})}\BibitemShut {NoStop}%
\bibitem [{\citenamefont {Wang}\ \emph {et~al.}(2024)\citenamefont {Wang}, \citenamefont {Chen}, \citenamefont {Lin}, \citenamefont {Hou}, \citenamefont {Lin},\ and\ \citenamefont {Chou}}]{PhysRevB.110.115154}%
  \BibitemOpen
  \bibfield  {author} {\bibinfo {author} {\bibfnamefont {W.-C.}\ \bibnamefont {Wang}}, \bibinfo {author} {\bibfnamefont {F.-W.}\ \bibnamefont {Chen}}, \bibinfo {author} {\bibfnamefont {K.-S.}\ \bibnamefont {Lin}}, \bibinfo {author} {\bibfnamefont {J.~T.}\ \bibnamefont {Hou}}, \bibinfo {author} {\bibfnamefont {H.-C.}\ \bibnamefont {Lin}},\ and\ \bibinfo {author} {\bibfnamefont {M.-Y.}\ \bibnamefont {Chou}},\ }\bibfield  {title} {\bibinfo {title} {Role of the fermi ring in determining the magic angles in twisted bilayer graphene},\ }\href {https://doi.org/10.1103/PhysRevB.110.115154} {\bibfield  {journal} {\bibinfo  {journal} {Phys. Rev. B}\ }\textbf {\bibinfo {volume} {110}},\ \bibinfo {pages} {115154} (\bibinfo {year} {2024})}\BibitemShut {NoStop}%
\bibitem [{\citenamefont {Carr}\ \emph {et~al.}(2020)\citenamefont {Carr}, \citenamefont {Li}, \citenamefont {Zhu}, \citenamefont {Kaxiras}, \citenamefont {Sachdev},\ and\ \citenamefont {Kruchkov}}]{Stephen_2020}%
  \BibitemOpen
  \bibfield  {author} {\bibinfo {author} {\bibfnamefont {S.}~\bibnamefont {Carr}}, \bibinfo {author} {\bibfnamefont {C.}~\bibnamefont {Li}}, \bibinfo {author} {\bibfnamefont {Z.}~\bibnamefont {Zhu}}, \bibinfo {author} {\bibfnamefont {E.}~\bibnamefont {Kaxiras}}, \bibinfo {author} {\bibfnamefont {S.}~\bibnamefont {Sachdev}},\ and\ \bibinfo {author} {\bibfnamefont {A.}~\bibnamefont {Kruchkov}},\ }\bibfield  {title} {\bibinfo {title} {Ultraheavy and ultrarelativistic dirac quasiparticles in sandwiched graphenes},\ }\href {https://doi.org/10.1021/acs.nanolett.9b04979} {\bibfield  {journal} {\bibinfo  {journal} {Nano Letters}\ }\textbf {\bibinfo {volume} {20}},\ \bibinfo {pages} {3030} (\bibinfo {year} {2020})}\BibitemShut {NoStop}%
\bibitem [{\citenamefont {Devakul}\ \emph {et~al.}(2023)\citenamefont {Devakul}, \citenamefont {Ledwith}, \citenamefont {Xia}, \citenamefont {Uri}, \citenamefont {de~la Barrera}, \citenamefont {Jarillo-Herrero},\ and\ \citenamefont {Fu}}]{Devakul2023}%
  \BibitemOpen
  \bibfield  {author} {\bibinfo {author} {\bibfnamefont {T.}~\bibnamefont {Devakul}}, \bibinfo {author} {\bibfnamefont {P.~J.}\ \bibnamefont {Ledwith}}, \bibinfo {author} {\bibfnamefont {L.-Q.}\ \bibnamefont {Xia}}, \bibinfo {author} {\bibfnamefont {A.}~\bibnamefont {Uri}}, \bibinfo {author} {\bibfnamefont {S.~C.}\ \bibnamefont {de~la Barrera}}, \bibinfo {author} {\bibfnamefont {P.}~\bibnamefont {Jarillo-Herrero}},\ and\ \bibinfo {author} {\bibfnamefont {L.}~\bibnamefont {Fu}},\ }\bibfield  {title} {\bibinfo {title} {Magic-angle helical trilayer graphene},\ }\href {https://doi.org/10.1126/sciadv.adi6063} {\bibfield  {journal} {\bibinfo  {journal} {Science Advances}\ }\textbf {\bibinfo {volume} {9}},\ \bibinfo {pages} {eadi6063} (\bibinfo {year} {2023})}\BibitemShut {NoStop}%
\bibitem [{\citenamefont {Ma}\ \emph {et~al.}(2021)\citenamefont {Ma}, \citenamefont {Li}, \citenamefont {Zheng}, \citenamefont {Xiao}, \citenamefont {Jiang}, \citenamefont {Gao},\ and\ \citenamefont {Xie}}]{theory2}%
  \BibitemOpen
  \bibfield  {author} {\bibinfo {author} {\bibfnamefont {Z.}~\bibnamefont {Ma}}, \bibinfo {author} {\bibfnamefont {S.}~\bibnamefont {Li}}, \bibinfo {author} {\bibfnamefont {Y.~W.}\ \bibnamefont {Zheng}}, \bibinfo {author} {\bibfnamefont {M.~M.}\ \bibnamefont {Xiao}}, \bibinfo {author} {\bibfnamefont {H.}~\bibnamefont {Jiang}}, \bibinfo {author} {\bibfnamefont {J.~H.}\ \bibnamefont {Gao}},\ and\ \bibinfo {author} {\bibfnamefont {X.~C.}\ \bibnamefont {Xie}},\ }\bibfield  {title} {\bibinfo {title} {Topological flat bands in twisted trilayer graphene},\ }\href {https://doi.org/10.1016/j.scib.2020.10.004} {\bibfield  {journal} {\bibinfo  {journal} {Science Bulletin}\ }\textbf {\bibinfo {volume} {66}},\ \bibinfo {pages} {18} (\bibinfo {year} {2021})}\BibitemShut {NoStop}%
\bibitem [{\citenamefont {Xu}\ \emph {et~al.}(2021)\citenamefont {Xu}, \citenamefont {Ezzi}, \citenamefont {Balakrishnan}, \citenamefont {Garcia-Ruiz}, \citenamefont {Tsim}, \citenamefont {Mullan}, \citenamefont {Barrier}, \citenamefont {Xin}, \citenamefont {Piot}, \citenamefont {Taniguchi}, \citenamefont {Watanabe}, \citenamefont {Carvalho}, \citenamefont {Mishchenko}, \citenamefont {Geim}, \citenamefont {Fal’ko}, \citenamefont {Adam}, \citenamefont {Neto}, \citenamefont {Novoselov},\ and\ \citenamefont {Shi}}]{Shuigang_2021}%
  \BibitemOpen
  \bibfield  {author} {\bibinfo {author} {\bibfnamefont {S.}~\bibnamefont {Xu}}, \bibinfo {author} {\bibfnamefont {M.~M.~A.}\ \bibnamefont {Ezzi}}, \bibinfo {author} {\bibfnamefont {N.}~\bibnamefont {Balakrishnan}}, \bibinfo {author} {\bibfnamefont {A.}~\bibnamefont {Garcia-Ruiz}}, \bibinfo {author} {\bibfnamefont {B.}~\bibnamefont {Tsim}}, \bibinfo {author} {\bibfnamefont {C.}~\bibnamefont {Mullan}}, \bibinfo {author} {\bibfnamefont {J.}~\bibnamefont {Barrier}}, \bibinfo {author} {\bibfnamefont {N.}~\bibnamefont {Xin}}, \bibinfo {author} {\bibfnamefont {B.~A.}\ \bibnamefont {Piot}}, \bibinfo {author} {\bibfnamefont {T.}~\bibnamefont {Taniguchi}}, \bibinfo {author} {\bibfnamefont {K.}~\bibnamefont {Watanabe}}, \bibinfo {author} {\bibfnamefont {A.}~\bibnamefont {Carvalho}}, \bibinfo {author} {\bibfnamefont {A.}~\bibnamefont {Mishchenko}}, \bibinfo {author} {\bibfnamefont {A.~K.}\ \bibnamefont {Geim}}, \bibinfo {author} {\bibfnamefont {V.~I.}\ \bibnamefont {Fal’ko}}, \bibinfo {author} {\bibfnamefont
  {S.}~\bibnamefont {Adam}}, \bibinfo {author} {\bibfnamefont {A.~H.~C.}\ \bibnamefont {Neto}}, \bibinfo {author} {\bibfnamefont {K.~S.}\ \bibnamefont {Novoselov}},\ and\ \bibinfo {author} {\bibfnamefont {Y.}~\bibnamefont {Shi}},\ }\bibfield  {title} {\bibinfo {title} {Tunable van hove singularities and correlated states in twisted monolayer–bilayer graphene},\ }\href {https://doi.org/10.1038/s41567-021-01172-9} {\bibfield  {journal} {\bibinfo  {journal} {Nature Physics}\ }\textbf {\bibinfo {volume} {17}},\ \bibinfo {pages} {619–626} (\bibinfo {year} {2021})}\BibitemShut {NoStop}%
\bibitem [{\citenamefont {Polshyn}\ \emph {et~al.}(2020)\citenamefont {Polshyn}, \citenamefont {Zhu}, \citenamefont {Kumar}, \citenamefont {Zhang}, \citenamefont {Yang}, \citenamefont {Tschirhart}, \citenamefont {Serlin}, \citenamefont {Watanabe}, \citenamefont {Taniguchi}, \citenamefont {MacDonald},\ and\ \citenamefont {Young}}]{H.Polshyn_2020}%
  \BibitemOpen
  \bibfield  {author} {\bibinfo {author} {\bibfnamefont {H.}~\bibnamefont {Polshyn}}, \bibinfo {author} {\bibfnamefont {J.}~\bibnamefont {Zhu}}, \bibinfo {author} {\bibfnamefont {M.~A.}\ \bibnamefont {Kumar}}, \bibinfo {author} {\bibfnamefont {Y.}~\bibnamefont {Zhang}}, \bibinfo {author} {\bibfnamefont {F.}~\bibnamefont {Yang}}, \bibinfo {author} {\bibfnamefont {C.~L.}\ \bibnamefont {Tschirhart}}, \bibinfo {author} {\bibfnamefont {M.}~\bibnamefont {Serlin}}, \bibinfo {author} {\bibfnamefont {K.}~\bibnamefont {Watanabe}}, \bibinfo {author} {\bibfnamefont {T.}~\bibnamefont {Taniguchi}}, \bibinfo {author} {\bibfnamefont {A.~H.}\ \bibnamefont {MacDonald}},\ and\ \bibinfo {author} {\bibfnamefont {A.~F.}\ \bibnamefont {Young}},\ }\bibfield  {title} {\bibinfo {title} {Electrical switching of magnetic order in an orbital chern insulator},\ }\href {https://doi.org/10.1038/s41586-020-2963-8} {\bibfield  {journal} {\bibinfo  {journal} {Nature}\ }\textbf {\bibinfo {volume} {588}},\ \bibinfo {pages} {66–70} (\bibinfo
  {year} {2020})}\BibitemShut {NoStop}%
\bibitem [{\citenamefont {Zhang}\ \emph {et~al.}(2023)\citenamefont {Zhang}, \citenamefont {Zhu}, \citenamefont {Soejima}, \citenamefont {Kahn}, \citenamefont {Watanabe}, \citenamefont {Taniguchi}, \citenamefont {Zettl}, \citenamefont {Wang}, \citenamefont {Zaletel},\ and\ \citenamefont {Crommie}}]{Canxun_2023}%
  \BibitemOpen
  \bibfield  {author} {\bibinfo {author} {\bibfnamefont {C.}~\bibnamefont {Zhang}}, \bibinfo {author} {\bibfnamefont {T.}~\bibnamefont {Zhu}}, \bibinfo {author} {\bibfnamefont {T.}~\bibnamefont {Soejima}}, \bibinfo {author} {\bibfnamefont {S.}~\bibnamefont {Kahn}}, \bibinfo {author} {\bibfnamefont {K.}~\bibnamefont {Watanabe}}, \bibinfo {author} {\bibfnamefont {T.}~\bibnamefont {Taniguchi}}, \bibinfo {author} {\bibfnamefont {A.}~\bibnamefont {Zettl}}, \bibinfo {author} {\bibfnamefont {F.}~\bibnamefont {Wang}}, \bibinfo {author} {\bibfnamefont {M.~P.}\ \bibnamefont {Zaletel}},\ and\ \bibinfo {author} {\bibfnamefont {M.~F.}\ \bibnamefont {Crommie}},\ }\bibfield  {title} {\bibinfo {title} {Local spectroscopy of a gate-switchable moiré quantum anomalous hall insulator},\ }\href {https://doi.org/10.1038/s41467-023-39110-3} {\bibfield  {journal} {\bibinfo  {journal} {Nature Communications}\ }\textbf {\bibinfo {volume} {14}},\ \bibinfo {pages} {3595} (\bibinfo {year} {2023})}\BibitemShut {NoStop}%
\bibitem [{\citenamefont {Polshyn}\ \emph {et~al.}(2022)\citenamefont {Polshyn}, \citenamefont {Zhang}, \citenamefont {Kumar}, \citenamefont {Soejima}, \citenamefont {Ledwith}, \citenamefont {Watanabe}, \citenamefont {Taniguchi}, \citenamefont {Vishwanath}, \citenamefont {Zaletel},\ and\ \citenamefont {Young}}]{H.Polshyn_2022}%
  \BibitemOpen
  \bibfield  {author} {\bibinfo {author} {\bibfnamefont {H.}~\bibnamefont {Polshyn}}, \bibinfo {author} {\bibfnamefont {Y.}~\bibnamefont {Zhang}}, \bibinfo {author} {\bibfnamefont {M.~A.}\ \bibnamefont {Kumar}}, \bibinfo {author} {\bibfnamefont {T.}~\bibnamefont {Soejima}}, \bibinfo {author} {\bibfnamefont {P.}~\bibnamefont {Ledwith}}, \bibinfo {author} {\bibfnamefont {K.}~\bibnamefont {Watanabe}}, \bibinfo {author} {\bibfnamefont {T.}~\bibnamefont {Taniguchi}}, \bibinfo {author} {\bibfnamefont {A.}~\bibnamefont {Vishwanath}}, \bibinfo {author} {\bibfnamefont {M.~P.}\ \bibnamefont {Zaletel}},\ and\ \bibinfo {author} {\bibfnamefont {A.~F.}\ \bibnamefont {Young}},\ }\bibfield  {title} {\bibinfo {title} {Topological charge density waves at half-integer filling of a moiré superlattice},\ }\href {https://doi.org/10.1038/s41567-021-01418-6} {\bibfield  {journal} {\bibinfo  {journal} {Nature Physics}\ }\textbf {\bibinfo {volume} {18}},\ \bibinfo {pages} {42–47} (\bibinfo {year} {2022})}\BibitemShut {NoStop}%
\bibitem [{\citenamefont {Chen}\ \emph {et~al.}(2021)\citenamefont {Chen}, \citenamefont {He}, \citenamefont {Zhang}, \citenamefont {Hsieh}, \citenamefont {Fei}, \citenamefont {Watanabe}, \citenamefont {Taniguchi}, \citenamefont {Cobden}, \citenamefont {Xu}, \citenamefont {Dean},\ and\ \citenamefont {Yankowitz}}]{Shaowen_2021}%
  \BibitemOpen
  \bibfield  {author} {\bibinfo {author} {\bibfnamefont {S.}~\bibnamefont {Chen}}, \bibinfo {author} {\bibfnamefont {M.}~\bibnamefont {He}}, \bibinfo {author} {\bibfnamefont {Y.~H.}\ \bibnamefont {Zhang}}, \bibinfo {author} {\bibfnamefont {V.}~\bibnamefont {Hsieh}}, \bibinfo {author} {\bibfnamefont {Z.}~\bibnamefont {Fei}}, \bibinfo {author} {\bibfnamefont {K.}~\bibnamefont {Watanabe}}, \bibinfo {author} {\bibfnamefont {T.}~\bibnamefont {Taniguchi}}, \bibinfo {author} {\bibfnamefont {D.~H.}\ \bibnamefont {Cobden}}, \bibinfo {author} {\bibfnamefont {X.}~\bibnamefont {Xu}}, \bibinfo {author} {\bibfnamefont {C.~R.}\ \bibnamefont {Dean}},\ and\ \bibinfo {author} {\bibfnamefont {M.}~\bibnamefont {Yankowitz}},\ }\bibfield  {title} {\bibinfo {title} {Electrically tunable correlated and topological states in twisted monolayer–bilayer graphene},\ }\href {https://doi.org/10.1038/s41567-020-01062-6} {\bibfield  {journal} {\bibinfo  {journal} {Nature Physics}\ }\textbf {\bibinfo {volume} {17}},\ \bibinfo {pages}
  {374–380} (\bibinfo {year} {2021})}\BibitemShut {NoStop}%
\bibitem [{\citenamefont {He}\ \emph {et~al.}(2021)\citenamefont {He}, \citenamefont {Zhang}, \citenamefont {Li}, \citenamefont {Fei}, \citenamefont {Watanabe}, \citenamefont {Taniguchi}, \citenamefont {Xu},\ and\ \citenamefont {Yankowitz}}]{Minhao_2021}%
  \BibitemOpen
  \bibfield  {author} {\bibinfo {author} {\bibfnamefont {M.}~\bibnamefont {He}}, \bibinfo {author} {\bibfnamefont {Y.~H.}\ \bibnamefont {Zhang}}, \bibinfo {author} {\bibfnamefont {Y.}~\bibnamefont {Li}}, \bibinfo {author} {\bibfnamefont {Z.}~\bibnamefont {Fei}}, \bibinfo {author} {\bibfnamefont {K.}~\bibnamefont {Watanabe}}, \bibinfo {author} {\bibfnamefont {T.}~\bibnamefont {Taniguchi}}, \bibinfo {author} {\bibfnamefont {X.}~\bibnamefont {Xu}},\ and\ \bibinfo {author} {\bibfnamefont {M.}~\bibnamefont {Yankowitz}},\ }\bibfield  {title} {\bibinfo {title} {Competing correlated states and abundant orbital magnetism in twisted monolayer-bilayer graphene},\ }\href {https://doi.org/10.1038/s41467-021-25044-1} {\bibfield  {journal} {\bibinfo  {journal} {Nature Communications}\ }\textbf {\bibinfo {volume} {12}},\ \bibinfo {pages} {4727} (\bibinfo {year} {2021})}\BibitemShut {NoStop}%
\bibitem [{\citenamefont {Li}\ \emph {et~al.}(2022)\citenamefont {Li}, \citenamefont {Wang}, \citenamefont {Xue}, \citenamefont {Wang}, \citenamefont {Zhang}, \citenamefont {Liu}, \citenamefont {Zhu}, \citenamefont {Watanabe}, \citenamefont {Taniguchi}, \citenamefont {Gao}, \citenamefont {Jiang},\ and\ \citenamefont {Mao}}]{Si_2022}%
  \BibitemOpen
  \bibfield  {author} {\bibinfo {author} {\bibfnamefont {S.~Y.}\ \bibnamefont {Li}}, \bibinfo {author} {\bibfnamefont {Z.}~\bibnamefont {Wang}}, \bibinfo {author} {\bibfnamefont {Y.}~\bibnamefont {Xue}}, \bibinfo {author} {\bibfnamefont {Y.}~\bibnamefont {Wang}}, \bibinfo {author} {\bibfnamefont {S.}~\bibnamefont {Zhang}}, \bibinfo {author} {\bibfnamefont {J.}~\bibnamefont {Liu}}, \bibinfo {author} {\bibfnamefont {Z.}~\bibnamefont {Zhu}}, \bibinfo {author} {\bibfnamefont {K.}~\bibnamefont {Watanabe}}, \bibinfo {author} {\bibfnamefont {T.}~\bibnamefont {Taniguchi}}, \bibinfo {author} {\bibfnamefont {H.~J.}\ \bibnamefont {Gao}}, \bibinfo {author} {\bibfnamefont {Y.}~\bibnamefont {Jiang}},\ and\ \bibinfo {author} {\bibfnamefont {J.}~\bibnamefont {Mao}},\ }\bibfield  {title} {\bibinfo {title} {Imaging topological and correlated insulating states in twisted monolayer-bilayer graphene},\ }\href {https://doi.org/10.1038/s41467-022-31851-x} {\bibfield  {journal} {\bibinfo  {journal} {Nature Communications}\ }\textbf
  {\bibinfo {volume} {13}},\ \bibinfo {pages} {4225} (\bibinfo {year} {2022})}\BibitemShut {NoStop}%
\bibitem [{\citenamefont {Nunn}\ \emph {et~al.}(2023)\citenamefont {Nunn}, \citenamefont {McEllistrim}, \citenamefont {Weston}, \citenamefont {Garcia-Ruiz}, \citenamefont {Watson}, \citenamefont {Mucha-Kruczynski}, \citenamefont {Cacho}, \citenamefont {Gorbachev}, \citenamefont {Fal’ko},\ and\ \citenamefont {Wilson}}]{Nunn_2023}%
  \BibitemOpen
  \bibfield  {author} {\bibinfo {author} {\bibfnamefont {J.~E.}\ \bibnamefont {Nunn}}, \bibinfo {author} {\bibfnamefont {A.}~\bibnamefont {McEllistrim}}, \bibinfo {author} {\bibfnamefont {A.}~\bibnamefont {Weston}}, \bibinfo {author} {\bibfnamefont {A.}~\bibnamefont {Garcia-Ruiz}}, \bibinfo {author} {\bibfnamefont {M.~D.}\ \bibnamefont {Watson}}, \bibinfo {author} {\bibfnamefont {M.}~\bibnamefont {Mucha-Kruczynski}}, \bibinfo {author} {\bibfnamefont {C.}~\bibnamefont {Cacho}}, \bibinfo {author} {\bibfnamefont {R.~V.}\ \bibnamefont {Gorbachev}}, \bibinfo {author} {\bibfnamefont {V.~I.}\ \bibnamefont {Fal’ko}},\ and\ \bibinfo {author} {\bibfnamefont {N.~R.}\ \bibnamefont {Wilson}},\ }\bibfield  {title} {\bibinfo {title} {Arpes signatures of few-layer twistronic graphenes},\ }\href {https://doi.org/10.1021/acs.nanolett.3c01173} {\bibfield  {journal} {\bibinfo  {journal} {Nano Letters}\ ,\ \bibinfo {pages} {5201–5208}} (\bibinfo {year} {2023})}\BibitemShut {NoStop}%
\bibitem [{\citenamefont {Zhang}\ \emph {et~al.}(2024)\citenamefont {Zhang}, \citenamefont {Li}, \citenamefont {Park}, \citenamefont {Jia}, \citenamefont {Chen}, \citenamefont {Li}, \citenamefont {Liu}, \citenamefont {Bao}, \citenamefont {Leconte}, \citenamefont {Zhou}, \citenamefont {Wang}, \citenamefont {Watanabe}, \citenamefont {Taniguchi}, \citenamefont {Avila}, \citenamefont {Dudin}, \citenamefont {Yu}, \citenamefont {Weng}, \citenamefont {Duan}, \citenamefont {Wu}, \citenamefont {Jung},\ and\ \citenamefont {Zhou}}]{zhang2024}%
  \BibitemOpen
  \bibfield  {author} {\bibinfo {author} {\bibfnamefont {H.}~\bibnamefont {Zhang}}, \bibinfo {author} {\bibfnamefont {Q.}~\bibnamefont {Li}}, \bibinfo {author} {\bibfnamefont {Y.}~\bibnamefont {Park}}, \bibinfo {author} {\bibfnamefont {Y.}~\bibnamefont {Jia}}, \bibinfo {author} {\bibfnamefont {W.}~\bibnamefont {Chen}}, \bibinfo {author} {\bibfnamefont {J.}~\bibnamefont {Li}}, \bibinfo {author} {\bibfnamefont {Q.}~\bibnamefont {Liu}}, \bibinfo {author} {\bibfnamefont {C.}~\bibnamefont {Bao}}, \bibinfo {author} {\bibfnamefont {N.}~\bibnamefont {Leconte}}, \bibinfo {author} {\bibfnamefont {S.}~\bibnamefont {Zhou}}, \bibinfo {author} {\bibfnamefont {Y.}~\bibnamefont {Wang}}, \bibinfo {author} {\bibfnamefont {K.}~\bibnamefont {Watanabe}}, \bibinfo {author} {\bibfnamefont {T.}~\bibnamefont {Taniguchi}}, \bibinfo {author} {\bibfnamefont {J.}~\bibnamefont {Avila}}, \bibinfo {author} {\bibfnamefont {P.}~\bibnamefont {Dudin}}, \bibinfo {author} {\bibfnamefont {P.}~\bibnamefont {Yu}}, \bibinfo {author} {\bibfnamefont
  {H.}~\bibnamefont {Weng}}, \bibinfo {author} {\bibfnamefont {W.}~\bibnamefont {Duan}}, \bibinfo {author} {\bibfnamefont {Q.}~\bibnamefont {Wu}}, \bibinfo {author} {\bibfnamefont {J.}~\bibnamefont {Jung}},\ and\ \bibinfo {author} {\bibfnamefont {S.}~\bibnamefont {Zhou}},\ }\bibfield  {title} {\bibinfo {title} {Observation of dichotomic field-tunable electronic structure in twisted monolayer-bilayer graphene},\ }\href {https://doi.org/10.1038/s41467-024-48166-8} {\bibfield  {journal} {\bibinfo  {journal} {Nature Communications}\ }\textbf {\bibinfo {volume} {15}},\ \bibinfo {pages} {3737} (\bibinfo {year} {2024})}\BibitemShut {NoStop}%
\bibitem [{\citenamefont {Liu}\ \emph {et~al.}(2019)\citenamefont {Liu}, \citenamefont {Ma}, \citenamefont {Gao},\ and\ \citenamefont {Dai}}]{Liu2019}%
  \BibitemOpen
  \bibfield  {author} {\bibinfo {author} {\bibfnamefont {J.}~\bibnamefont {Liu}}, \bibinfo {author} {\bibfnamefont {Z.}~\bibnamefont {Ma}}, \bibinfo {author} {\bibfnamefont {J.}~\bibnamefont {Gao}},\ and\ \bibinfo {author} {\bibfnamefont {X.}~\bibnamefont {Dai}},\ }\bibfield  {title} {\bibinfo {title} {Quantum valley hall effect, orbital magnetism, and anomalous hall effect in twisted multilayer graphene systems},\ }\href {https://doi.org/10.1103/PhysRevX.9.031021} {\bibfield  {journal} {\bibinfo  {journal} {Phys. Rev. X}\ }\textbf {\bibinfo {volume} {9}},\ \bibinfo {pages} {031021} (\bibinfo {year} {2019})}\BibitemShut {NoStop}%
\bibitem [{\citenamefont {Park}\ \emph {et~al.}(2020)\citenamefont {Park}, \citenamefont {Chittari},\ and\ \citenamefont {Jung}}]{theory1}%
  \BibitemOpen
  \bibfield  {author} {\bibinfo {author} {\bibfnamefont {Y.}~\bibnamefont {Park}}, \bibinfo {author} {\bibfnamefont {B.~L.}\ \bibnamefont {Chittari}},\ and\ \bibinfo {author} {\bibfnamefont {J.}~\bibnamefont {Jung}},\ }\bibfield  {title} {\bibinfo {title} {Gate-tunable topological flat bands in twisted monolayer-bilayer graphene},\ }\href {https://doi.org/10.1103/PhysRevB.102.035411} {\bibfield  {journal} {\bibinfo  {journal} {Physical Review B}\ }\textbf {\bibinfo {volume} {102}},\ \bibinfo {pages} {035411} (\bibinfo {year} {2020})}\BibitemShut {NoStop}%
\bibitem [{\citenamefont {Rademaker}\ \emph {et~al.}(2020)\citenamefont {Rademaker}, \citenamefont {Protopopov},\ and\ \citenamefont {Abanin}}]{theory3}%
  \BibitemOpen
  \bibfield  {author} {\bibinfo {author} {\bibfnamefont {L.}~\bibnamefont {Rademaker}}, \bibinfo {author} {\bibfnamefont {I.~V.}\ \bibnamefont {Protopopov}},\ and\ \bibinfo {author} {\bibfnamefont {D.~A.}\ \bibnamefont {Abanin}},\ }\bibfield  {title} {\bibinfo {title} {Topological flat bands and correlated states in twisted monolayer-bilayer graphene},\ }\href {https://doi.org/10.1103/PhysRevResearch.2.033150} {\bibfield  {journal} {\bibinfo  {journal} {Physical Review Research}\ }\textbf {\bibinfo {volume} {2}},\ \bibinfo {pages} {033150} (\bibinfo {year} {2020})}\BibitemShut {NoStop}%
\bibitem [{\citenamefont {Zhang}\ \emph {et~al.}(2022)\citenamefont {Zhang}, \citenamefont {Dai},\ and\ \citenamefont {Liu}}]{ZhangShihao2022}%
  \BibitemOpen
  \bibfield  {author} {\bibinfo {author} {\bibfnamefont {S.}~\bibnamefont {Zhang}}, \bibinfo {author} {\bibfnamefont {X.}~\bibnamefont {Dai}},\ and\ \bibinfo {author} {\bibfnamefont {J.}~\bibnamefont {Liu}},\ }\bibfield  {title} {\bibinfo {title} {Spin-polarized nematic order, quantum valley hall states, and field-tunable topological transitions in twisted multilayer graphene systems},\ }\href {https://doi.org/10.1103/PhysRevLett.128.026403} {\bibfield  {journal} {\bibinfo  {journal} {Phys. Rev. Lett.}\ }\textbf {\bibinfo {volume} {128}},\ \bibinfo {pages} {026403} (\bibinfo {year} {2022})}\BibitemShut {NoStop}%
\bibitem [{\citenamefont {Fang}\ and\ \citenamefont {Kaxiras}(2016)}]{TB_parameter_1}%
  \BibitemOpen
  \bibfield  {author} {\bibinfo {author} {\bibfnamefont {S.}~\bibnamefont {Fang}}\ and\ \bibinfo {author} {\bibfnamefont {E.}~\bibnamefont {Kaxiras}},\ }\bibfield  {title} {\bibinfo {title} {Electronic structure theory of weakly interacting bilayers},\ }\href {https://doi.org/10.1103/PhysRevB.93.235153} {\bibfield  {journal} {\bibinfo  {journal} {Physical Review B}\ }\textbf {\bibinfo {volume} {93}},\ \bibinfo {pages} {235153} (\bibinfo {year} {2016})}\BibitemShut {NoStop}%
\bibitem [{\citenamefont {Kresse}\ and\ \citenamefont {Furthmüller}(1996{\natexlab{a}})}]{Kresse1996}%
  \BibitemOpen
  \bibfield  {author} {\bibinfo {author} {\bibfnamefont {G.}~\bibnamefont {Kresse}}\ and\ \bibinfo {author} {\bibfnamefont {J.}~\bibnamefont {Furthmüller}},\ }\bibfield  {title} {\bibinfo {title} {Efficient iterative schemes for ab initio total-energy calculations using a plane-wave basis set},\ }\href {https://doi.org/10.1103/PhysRevB.54.11169} {\bibfield  {journal} {\bibinfo  {journal} {Physical Review B - Condensed Matter and Materials Physics}\ }\textbf {\bibinfo {volume} {54}},\ \bibinfo {pages} {11169} (\bibinfo {year} {1996}{\natexlab{a}})}\BibitemShut {NoStop}%
\bibitem [{\citenamefont {Kresse}\ and\ \citenamefont {Furthmüller}(1996{\natexlab{b}})}]{Kresse1996-2}%
  \BibitemOpen
  \bibfield  {author} {\bibinfo {author} {\bibfnamefont {G.}~\bibnamefont {Kresse}}\ and\ \bibinfo {author} {\bibfnamefont {J.}~\bibnamefont {Furthmüller}},\ }\bibfield  {title} {\bibinfo {title} {Efficiency of ab-initio total energy calculations for metals and semiconductors using a plane-wave basis set},\ }\href {https://doi.org/10.1016/0927-0256(96)00008-0} {\bibfield  {journal} {\bibinfo  {journal} {Computational Materials Science}\ }\textbf {\bibinfo {volume} {6}},\ \bibinfo {pages} {15} (\bibinfo {year} {1996}{\natexlab{b}})}\BibitemShut {NoStop}%
\bibitem [{\citenamefont {Blöchl}(1994)}]{Blöchl1994}%
  \BibitemOpen
  \bibfield  {author} {\bibinfo {author} {\bibfnamefont {P.~E.}\ \bibnamefont {Blöchl}},\ }\bibfield  {title} {\bibinfo {title} {Projector augmented-wave method},\ }\href {https://doi.org/10.1103/PhysRevB.50.17953} {\bibfield  {journal} {\bibinfo  {journal} {Physical Review B}\ }\textbf {\bibinfo {volume} {50}},\ \bibinfo {pages} {17953} (\bibinfo {year} {1994})}\BibitemShut {NoStop}%
\bibitem [{\citenamefont {Joubert}(1999)}]{Joubert1999}%
  \BibitemOpen
  \bibfield  {author} {\bibinfo {author} {\bibfnamefont {D.}~\bibnamefont {Joubert}},\ }\bibfield  {title} {\bibinfo {title} {From ultrasoft pseudopotentials to the projector augmented-wave method},\ }\href {https://doi.org/10.1103/PhysRevB.59.1758} {\bibfield  {journal} {\bibinfo  {journal} {Physical Review B - Condensed Matter and Materials Physics}\ }\textbf {\bibinfo {volume} {59}},\ \bibinfo {pages} {1758} (\bibinfo {year} {1999})}\BibitemShut {NoStop}%
\bibitem [{\citenamefont {Perdew}\ \emph {et~al.}(1996)\citenamefont {Perdew}, \citenamefont {Burke},\ and\ \citenamefont {Ernzerhof}}]{Perdew1996}%
  \BibitemOpen
  \bibfield  {author} {\bibinfo {author} {\bibfnamefont {J.~P.}\ \bibnamefont {Perdew}}, \bibinfo {author} {\bibfnamefont {K.}~\bibnamefont {Burke}},\ and\ \bibinfo {author} {\bibfnamefont {M.}~\bibnamefont {Ernzerhof}},\ }\bibfield  {title} {\bibinfo {title} {Generalized gradient approximation made simple},\ }\href {https://doi.org/10.1103/PhysRevLett.77.3865} {\bibfield  {journal} {\bibinfo  {journal} {Physical Review Letters}\ }\textbf {\bibinfo {volume} {77}},\ \bibinfo {pages} {3865} (\bibinfo {year} {1996})}\BibitemShut {NoStop}%
\bibitem [{\citenamefont {King-Smith}\ and\ \citenamefont {Vanderbilt}(1993)}]{King-Smith1993}%
  \BibitemOpen
  \bibfield  {author} {\bibinfo {author} {\bibfnamefont {R.~D.}\ \bibnamefont {King-Smith}}\ and\ \bibinfo {author} {\bibfnamefont {D.}~\bibnamefont {Vanderbilt}},\ }\bibfield  {title} {\bibinfo {title} {Theory of polarization of crystalline solids},\ }\href {https://doi.org/10.1103/PhysRevB.47.1651} {\bibfield  {journal} {\bibinfo  {journal} {Physical Review B}\ }\textbf {\bibinfo {volume} {47}},\ \bibinfo {pages} {1651} (\bibinfo {year} {1993})}\BibitemShut {NoStop}%
\bibitem [{\citenamefont {Morell}\ \emph {et~al.}(2013)\citenamefont {Morell}, \citenamefont {Pacheco}, \citenamefont {Chico},\ and\ \citenamefont {Brey}}]{tight-binding}%
  \BibitemOpen
  \bibfield  {author} {\bibinfo {author} {\bibfnamefont {E.~S.}\ \bibnamefont {Morell}}, \bibinfo {author} {\bibfnamefont {M.}~\bibnamefont {Pacheco}}, \bibinfo {author} {\bibfnamefont {L.}~\bibnamefont {Chico}},\ and\ \bibinfo {author} {\bibfnamefont {L.}~\bibnamefont {Brey}},\ }\bibfield  {title} {\bibinfo {title} {Electronic properties of twisted trilayer graphene},\ }\href {https://doi.org/10.1103/PhysRevB.87.125414} {\bibfield  {journal} {\bibinfo  {journal} {Physical Review B - Condensed Matter and Materials Physics}\ }\textbf {\bibinfo {volume} {87}},\ \bibinfo {pages} {125414} (\bibinfo {year} {2013})}\BibitemShut {NoStop}%
\bibitem [{\citenamefont {Rickhaus}\ \emph {et~al.}(2019)\citenamefont {Rickhaus}, \citenamefont {Zheng}, \citenamefont {Lado}, \citenamefont {Lee}, \citenamefont {Kurzmann}, \citenamefont {Eich}, \citenamefont {Pisoni}, \citenamefont {Tong}, \citenamefont {Garreis}, \citenamefont {Gold}, \citenamefont {Masseroni}, \citenamefont {Taniguchi}, \citenamefont {Wantanabe}, \citenamefont {Ihn},\ and\ \citenamefont {Ensslin}}]{TDBG1}%
  \BibitemOpen
  \bibfield  {author} {\bibinfo {author} {\bibfnamefont {P.}~\bibnamefont {Rickhaus}}, \bibinfo {author} {\bibfnamefont {G.}~\bibnamefont {Zheng}}, \bibinfo {author} {\bibfnamefont {J.~L.}\ \bibnamefont {Lado}}, \bibinfo {author} {\bibfnamefont {Y.}~\bibnamefont {Lee}}, \bibinfo {author} {\bibfnamefont {A.}~\bibnamefont {Kurzmann}}, \bibinfo {author} {\bibfnamefont {M.}~\bibnamefont {Eich}}, \bibinfo {author} {\bibfnamefont {R.}~\bibnamefont {Pisoni}}, \bibinfo {author} {\bibfnamefont {C.}~\bibnamefont {Tong}}, \bibinfo {author} {\bibfnamefont {R.}~\bibnamefont {Garreis}}, \bibinfo {author} {\bibfnamefont {C.}~\bibnamefont {Gold}}, \bibinfo {author} {\bibfnamefont {M.}~\bibnamefont {Masseroni}}, \bibinfo {author} {\bibfnamefont {T.}~\bibnamefont {Taniguchi}}, \bibinfo {author} {\bibfnamefont {K.}~\bibnamefont {Wantanabe}}, \bibinfo {author} {\bibfnamefont {T.}~\bibnamefont {Ihn}},\ and\ \bibinfo {author} {\bibfnamefont {K.}~\bibnamefont {Ensslin}},\ }\bibfield  {title} {\bibinfo {title} {Gap opening in
  twisted double bilayer graphene by crystal fields},\ }\href {https://doi.org/10.1021/acs.nanolett.9b03660} {\bibfield  {journal} {\bibinfo  {journal} {Nano Letters}\ }\textbf {\bibinfo {volume} {19}},\ \bibinfo {pages} {8821–8828} (\bibinfo {year} {2019})}\BibitemShut {NoStop}%
\bibitem [{\citenamefont {Culchac}\ \emph {et~al.}(2020)\citenamefont {Culchac}, \citenamefont {Grande}, \citenamefont {Capaz}, \citenamefont {Chico},\ and\ \citenamefont {Morell}}]{TDBG2}%
  \BibitemOpen
  \bibfield  {author} {\bibinfo {author} {\bibfnamefont {F.~J.}\ \bibnamefont {Culchac}}, \bibinfo {author} {\bibfnamefont {R.~R.~D.}\ \bibnamefont {Grande}}, \bibinfo {author} {\bibfnamefont {R.~B.}\ \bibnamefont {Capaz}}, \bibinfo {author} {\bibfnamefont {L.}~\bibnamefont {Chico}},\ and\ \bibinfo {author} {\bibfnamefont {E.~S.}\ \bibnamefont {Morell}},\ }\bibfield  {title} {\bibinfo {title} {Flat bands and gaps in twisted double bilayer graphene},\ }\href {https://doi.org/10.1039/c9nr10830k} {\bibfield  {journal} {\bibinfo  {journal} {Nanoscale}\ }\textbf {\bibinfo {volume} {12}},\ \bibinfo {pages} {5014} (\bibinfo {year} {2020})}\BibitemShut {NoStop}%
\bibitem [{\citenamefont {Haddadi}\ \emph {et~al.}(2020)\citenamefont {Haddadi}, \citenamefont {Wu}, \citenamefont {Kruchkov},\ and\ \citenamefont {Yazyev}}]{TDBG3}%
  \BibitemOpen
  \bibfield  {author} {\bibinfo {author} {\bibfnamefont {F.}~\bibnamefont {Haddadi}}, \bibinfo {author} {\bibfnamefont {Q.~S.}\ \bibnamefont {Wu}}, \bibinfo {author} {\bibfnamefont {A.~J.}\ \bibnamefont {Kruchkov}},\ and\ \bibinfo {author} {\bibfnamefont {O.~V.}\ \bibnamefont {Yazyev}},\ }\bibfield  {title} {\bibinfo {title} {Moiré flat bands in twisted double bilayer graphene},\ }\href {https://doi.org/10.1021/acs.nanolett.9b05117} {\bibfield  {journal} {\bibinfo  {journal} {Nano Letters}\ }\textbf {\bibinfo {volume} {20}},\ \bibinfo {pages} {2410–2415} (\bibinfo {year} {2020})}\BibitemShut {NoStop}%
\bibitem [{\citenamefont {Li}\ \emph {et~al.}(2010)\citenamefont {Li}, \citenamefont {Luican}, \citenamefont {Santos}, \citenamefont {Neto}, \citenamefont {Reina}, \citenamefont {Kong},\ and\ \citenamefont {Andrei}}]{TBG_localization_1}%
  \BibitemOpen
  \bibfield  {author} {\bibinfo {author} {\bibfnamefont {G.}~\bibnamefont {Li}}, \bibinfo {author} {\bibfnamefont {A.}~\bibnamefont {Luican}}, \bibinfo {author} {\bibfnamefont {J.~M. L.~D.}\ \bibnamefont {Santos}}, \bibinfo {author} {\bibfnamefont {A.~H.~C.}\ \bibnamefont {Neto}}, \bibinfo {author} {\bibfnamefont {A.}~\bibnamefont {Reina}}, \bibinfo {author} {\bibfnamefont {J.}~\bibnamefont {Kong}},\ and\ \bibinfo {author} {\bibfnamefont {E.~Y.}\ \bibnamefont {Andrei}},\ }\bibfield  {title} {\bibinfo {title} {Observation of van hove singularities in twisted graphene layers},\ }\href {https://doi.org/10.1038/nphys1463} {\bibfield  {journal} {\bibinfo  {journal} {Nature Physics}\ }\textbf {\bibinfo {volume} {6}},\ \bibinfo {pages} {109–113} (\bibinfo {year} {2010})}\BibitemShut {NoStop}%
\bibitem [{\citenamefont {Brihuega}\ \emph {et~al.}(2012)\citenamefont {Brihuega}, \citenamefont {Mallet}, \citenamefont {González-Herrero}, \citenamefont {Laissardière}, \citenamefont {Ugeda}, \citenamefont {Magaud}, \citenamefont {Gómez-Rodríguez}, \citenamefont {Ynduráin},\ and\ \citenamefont {Veuillen}}]{TBG_localization_2}%
  \BibitemOpen
  \bibfield  {author} {\bibinfo {author} {\bibfnamefont {I.}~\bibnamefont {Brihuega}}, \bibinfo {author} {\bibfnamefont {P.}~\bibnamefont {Mallet}}, \bibinfo {author} {\bibfnamefont {H.}~\bibnamefont {González-Herrero}}, \bibinfo {author} {\bibfnamefont {G.~T.~D.}\ \bibnamefont {Laissardière}}, \bibinfo {author} {\bibfnamefont {M.~M.}\ \bibnamefont {Ugeda}}, \bibinfo {author} {\bibfnamefont {L.}~\bibnamefont {Magaud}}, \bibinfo {author} {\bibfnamefont {J.~M.}\ \bibnamefont {Gómez-Rodríguez}}, \bibinfo {author} {\bibfnamefont {F.}~\bibnamefont {Ynduráin}},\ and\ \bibinfo {author} {\bibfnamefont {J.~Y.}\ \bibnamefont {Veuillen}},\ }\bibfield  {title} {\bibinfo {title} {Unraveling the intrinsic and robust nature of van hove singularities in twisted bilayer graphene by scanning tunneling microscopy and theoretical analysis},\ }\href {https://doi.org/10.1103/PhysRevLett.109.196802} {\bibfield  {journal} {\bibinfo  {journal} {Physical Review Letters}\ }\textbf {\bibinfo {volume} {109}},\ \bibinfo {pages}
  {196802} (\bibinfo {year} {2012})}\BibitemShut {NoStop}%
\bibitem [{\citenamefont {Kerelsky}\ \emph {et~al.}(2019)\citenamefont {Kerelsky}, \citenamefont {McGilly}, \citenamefont {Kennes}, \citenamefont {Xian}, \citenamefont {Yankowitz}, \citenamefont {Chen}, \citenamefont {Watanabe}, \citenamefont {Taniguchi}, \citenamefont {Hone}, \citenamefont {Dean}, \citenamefont {Rubio},\ and\ \citenamefont {Pasupathy}}]{TBG_localization_3}%
  \BibitemOpen
  \bibfield  {author} {\bibinfo {author} {\bibfnamefont {A.}~\bibnamefont {Kerelsky}}, \bibinfo {author} {\bibfnamefont {L.~J.}\ \bibnamefont {McGilly}}, \bibinfo {author} {\bibfnamefont {D.~M.}\ \bibnamefont {Kennes}}, \bibinfo {author} {\bibfnamefont {L.}~\bibnamefont {Xian}}, \bibinfo {author} {\bibfnamefont {M.}~\bibnamefont {Yankowitz}}, \bibinfo {author} {\bibfnamefont {S.}~\bibnamefont {Chen}}, \bibinfo {author} {\bibfnamefont {K.}~\bibnamefont {Watanabe}}, \bibinfo {author} {\bibfnamefont {T.}~\bibnamefont {Taniguchi}}, \bibinfo {author} {\bibfnamefont {J.}~\bibnamefont {Hone}}, \bibinfo {author} {\bibfnamefont {C.}~\bibnamefont {Dean}}, \bibinfo {author} {\bibfnamefont {A.}~\bibnamefont {Rubio}},\ and\ \bibinfo {author} {\bibfnamefont {A.~N.}\ \bibnamefont {Pasupathy}},\ }\bibfield  {title} {\bibinfo {title} {Maximized electron interactions at the magic angle in twisted bilayer graphene},\ }\href {https://doi.org/10.1038/s41586-019-1431-9} {\bibfield  {journal} {\bibinfo  {journal} {Nature}\ }\textbf
  {\bibinfo {volume} {572}},\ \bibinfo {pages} {95–100} (\bibinfo {year} {2019})}\BibitemShut {NoStop}%
\bibitem [{\citenamefont {Tong}\ \emph {et~al.}(2022)\citenamefont {Tong}, \citenamefont {Tong}, \citenamefont {Yang}, \citenamefont {Zhou}, \citenamefont {Wu}, \citenamefont {Tian}, \citenamefont {Zhang}, \citenamefont {Zhang}, \citenamefont {Qin},\ and\ \citenamefont {Yin}}]{Ling_2022}%
  \BibitemOpen
  \bibfield  {author} {\bibinfo {author} {\bibfnamefont {L.~H.}\ \bibnamefont {Tong}}, \bibinfo {author} {\bibfnamefont {Q.}~\bibnamefont {Tong}}, \bibinfo {author} {\bibfnamefont {L.~Z.}\ \bibnamefont {Yang}}, \bibinfo {author} {\bibfnamefont {Y.~Y.}\ \bibnamefont {Zhou}}, \bibinfo {author} {\bibfnamefont {Q.}~\bibnamefont {Wu}}, \bibinfo {author} {\bibfnamefont {Y.}~\bibnamefont {Tian}}, \bibinfo {author} {\bibfnamefont {L.}~\bibnamefont {Zhang}}, \bibinfo {author} {\bibfnamefont {L.}~\bibnamefont {Zhang}}, \bibinfo {author} {\bibfnamefont {Z.}~\bibnamefont {Qin}},\ and\ \bibinfo {author} {\bibfnamefont {L.~J.}\ \bibnamefont {Yin}},\ }\bibfield  {title} {\bibinfo {title} {Spectroscopic visualization of flat bands in magic-angle twisted monolayer-bilayer graphene: Coexistence of localization and delocalization},\ }\href {https://doi.org/10.1103/PhysRevLett.128.126401} {\bibfield  {journal} {\bibinfo  {journal} {Physical Review Letters}\ }\textbf {\bibinfo {volume} {128}},\ \bibinfo {pages} {126401} (\bibinfo
  {year} {2022})}\BibitemShut {NoStop}%
\end{thebibliography}%

\end{document}